\documentclass[aps,pra,twocolumn,showpacs,floatfix]{revtex4-1}

\usepackage{epsfig,amsmath,amssymb}
\usepackage{graphics} 
\usepackage{subfigure}
\usepackage{color}
\usepackage{multirow}
\usepackage{mathrsfs}
\usepackage{natbib}

\begin{document}

\title
{
Frustrated Heisenberg antiferromagnet on the honeycomb lattice: Spin gap and low-energy parameters
}

\author
{R.~F.~Bishop and P.~H.~Y.~Li}
\affiliation
{School of Physics and Astronomy, Schuster Building, The University of Manchester, Manchester, M13 9PL, UK}

\author
{O.~G\"otze and J.~Richter}
\affiliation
{Institut f\"ur Theoretische Physik, Otto-von-Guericke Universit\"at Magdeburg, 39016 Magdeburg, Germany}

\author
{C.~E.~Campbell}
\affiliation
{School of Physics and Astronomy, University of Minnesota, 116 Church Street SE, Minneapolis, Minnesota 55455, USA}


\begin{abstract}
  We use the coupled cluster method implemented to high orders
  of approximation to investigate the frustrated spin-$\frac{1}{2}$
  $J_{1}$--$J_{2}$--$J_{3}$ antiferromagnet on the honeycomb lattice
  with isotropic Heisenberg interactions of strength $J_{1} > 0$
  between nearest-neighbor pairs, $J_{2}>0$ between next-nearest
  neighbor pairs, and $J_{3}>0$ between next-next-neareast-neighbor
  pairs of spins.  In particular, we study both the ground-state (GS) and
  lowest-lying triplet excited-state properties in the case
  $J_{3}=J_{2} \equiv \kappa J_{1}$, in the window $0 \leq \kappa \leq
  1$ of the frustration parameter, which includes the (tricritical)
  point of maximum classical frustration at $\kappa_{{\rm cl}} = \frac{1}{2}$.  We present GS results for
  the spin stiffness, $\rho_{s}$, and the zero-field uniform magnetic
  susceptibility, $\chi$, which complement our earlier results for the
  GS energy per spin, $E/N$, and staggered magnetization, $M$, to
  yield a complete set of accurate low-energy parameters for the
  model.  Our results all point towards a phase diagram containing two
  quasiclassical antiferromagnetic phases, one with N\'{e}el order for
  $\kappa < \kappa_{c_{1}}$, and the other with collinear striped
  order for $\kappa > \kappa_{c_{2}}$.  The results for both $\chi$
  and the spin gap $\Delta$ provide compelling evidence for a
  disordered quantum paramagnetic phase that is gapped over a
  considerable portion of the intermediate region $\kappa_{c_{1}} <
  \kappa < \kappa_{c_{2}}$, especially close to the two quantum
  critical points at $\kappa_{c_{1}}$ and $\kappa_{c_{2}}$.  Each of
  our fully independent sets of results for the low-energy parameters
  is consistent with the values $\kappa_{c_{1}} = 0.45 \pm
  0.02$ and $\kappa_{c_{2}} = 0.60 \pm 0.02$, and with the transition
  at $\kappa_{c_{1}}$ being of continuous (and hence probably of the
  deconfined) type and that at $\kappa_{c_{2}}$ being of first-order
  type.
\end{abstract}


\maketitle

\section{INTRODUCTION}
\label{introd_sec}
The roles played by quantum fluctuations and frustration on the
ordering properties of the ground-state (GS) phases of Heisenberg
systems of interacting spins placed on the sites of a regular periodic
lattice continue to be the subject of intense study.  Other things
being equal, quantum fluctuations tend to be larger for lower spatial
dimensionality $D$, lower values of the spin quantum number $s$, and
lower values of the lattice coordination number $z$.  The
Mermin-Wagner theorem \cite{Mermin:1966} rules out the possibility of
magnetic order in an isotropic Heisenberg system when $D=1$ even at
zero temperature ($T=0$), since it is not possible to break a
continuous symmetry in a 1D system even at $T=0$.  Similarly, the
Mermin-Wagner theorem also implies the absence of magnetic order in
any isotropic 2D Heisenberg model at all nonzero temperatures ($T>0$).
Thus, the study of 2D spin-lattice systems at $T=0$ occupies a very
special role for the study of quantum phase transitions.  In order to
maximize the role of quantum fluctuations it is then natural to focus
special attention on spin-$\frac{1}{2}$ systems on the honeycomb
lattice, which have the lowest values of both $s$ and $z$ ($=3$ in
this case).

The simplest archetypal such model is perhaps then the pure isotropic
Heisenberg model, which comprises antiferromagnetic Heisenberg bonds only
between nearest-neighbor (NN) pairs on the lattice, all with
identical exchange coupling strength $J_{1}>0$.  This model has been
studied many times in the past, using a variety of theoretical tools,
including, for example, the exact diagonalization (ED) of small
finite-sized lattices \cite{Oitmaa:1978_honey,Richter:2004_triang_ED},
spin-wave theory (SWT) \cite{Zheng:1991_honey}, Schwinger-boson
mean-field theory (SBMFT) \cite{Mattsson:1994_honey}, a linked-cluster
series expansion (SE) around the Ising limit \cite{Oitmaa:1992_honey},
quantum Monte Carlo (QMC) simulations
\cite{Reger:1989_honey,Castra:2006_honey,Jiang:2008_honey,Low:2009_honey,Jiang:2012_honey},
and the coupled cluster method (CCM)
\cite{Bishop:1998_honey,DJJFarnell:2014_archimedeanLatt}.  From many such studies there
is now a broad consensus that the N\'{e}el order of the classical ($s
\rightarrow \infty$) Heisenberg antiferromagnet (HAF) on a bipartite
lattice at $T=0$ is not destroyed by quantum fluctuations in the
$s=\frac{1}{2}$ case for the honeycomb lattice, albeit with a much
reduced value for the N\'{e}el order parameter.  Thus, the staggered
(or sublattice) magnetization $M$ for the spin-$\frac{1}{2}$
honeycomb-lattice HAF is generally agreed to take a value of about 54\%
of the classical value.  The role of reducing the lattice coordination
number, for example, from $z=4$ for the square lattice to $z=3$ for
the honeycomb lattice can be seen from the corresponding value for $M$
for the spin-$\frac{1}{2}$ square-lattice HAF, which is generally
agreed to be about 62\% of the classical value.

Another fundamental difference between the square and honeycomb
lattices is that, although both are bipartite, whereas the former is a
Bravais lattice, the latter is not.  Instead, the honeycomb lattice
has two sites per unit cell and comprises two interlocking triangular
Bravais lattices.  Thus, the translational invariance of the full
honeycomb lattice is broken for any type of state in general.  This
has the consequence, for example, that the transition from magnetic
disorder at $T>0$ to N\'{e}el antiferromagnetic (AFM) order at $T=0$
is not accompanied by a reduction in the spatial symmetry for the
honeycomb lattice, whereas it is for the square lattice.  For the
spin-$\frac{1}{2}$ square-lattice HAF, with one site per unit cell,
the Lieb-Schultz-Mattis-Hastings theorem
\cite{Lieb:1961_LSMH-theorem,Hastings:2004_Lieb-LSM-Hast-theorem} then
applies.  The theorem broadly asserts that a system with
half-odd-integer spin in the unit cell cannot have a gap and a unique
ground state.  Thus, for the square lattice, any gapped state must be
accompanied by a symmetry breaking.  By contrast, for a lattice, such
as the honeycomb lattice, with an even number of sites per unit cell,
the generalization by Hastings
\cite{Hastings:2004_Lieb-LSM-Hast-theorem} of the Lieb-Schultz-Mattis
theorem \cite{Lieb:1961_LSMH-theorem} does not apply.  In the case of
spin-$\frac{1}{2}$ models on the honeycomb lattice, unlike on the
square lattice, one can in principle have a gapped magnetically disordered GS
phase that does not break any symmetry, and which has only trivial
topological properties.

It is also interesting to note that several spin-$\frac{1}{2}$
honeycomb-lattice systems have been realized experimentally.  For
example, recent calculations \cite{Tsirlin:2010_honey} of the
low-dimensional magnetic material $\beta$-Cu$_{2}$V$_{2}$O$_{7}$ have
shown that its magnetic properties can be described by a
spin-$\frac{1}{2}$ anisotropic honeycomb HAF model, albeit with two
inequivalent NN bonds arising from the anisotropic exchange
interactions.  Another example is the compound
Na$_{3}$Cu$_{2}$SbO$_{6}$ in which the apparently hexagonal
arrangement of the spin-$\frac{1}{2}$ Cu$^{2+}$ ions in the copper
oxide layers has been taken as evidence of honeycomb-lattice magnetism
\cite{Miura:2006_honey}.  However, it has also been pointed out
\cite{Tsirlin:2010_honey} that the structural distortion of the
lattice and the orbital states of the Cu ions again conspire to make
different NN AFM bonds on the lattice sufficiently inequivalent as to
induce 1D-type magnetic behavior.  Another compound with
spin-$\frac{1}{2}$ Cu$^{2+}$ ions arranged in a honeycomb lattice in
the copper oxide layers is {I}n{C}u$_{2/3}${V}$_{1/3}${O}$_{3}$
\cite{Kataev:2005_honey,Okubo:2011_honey}.  This is probably the only
known substance described by a spin-$\frac{1}{2}$ HAF on the honeycomb
lattice with equivalent NN exchange interactions.  Nevertheless, even
here decisive comparison between experiment and theory is made
difficult by the tendency of the material to disorder structurally,
due to the mixing of the magnetic spin-$\frac{1}{2}$ Cu$^{2+}$ ions
with the nonmagnetic V$^{5+}$ ions
\cite{Moller:2008_honey,Yehia:2010_honey}.

Although the N\'{e}el order in the spin-$\frac{1}{2}$ HAF on the
honeycomb lattice with AFM bonds on NN sites only, all with the same
strength $J_{1}$, is stable against quantum fluctuations, the much
reduced value of the staggered magnetization order parameter from its
classical value suggests that it is likely to be rather fragile
against the onset of frustrating interactions.  In recent years,
therefore, it has become of great interest to investigate the
corresponding model where the NN bonds with strength $J_{1}>0$ are
augmented by frustrating next-nearest-neighbor (NNN) bonds with
strength $J_{2}>0$, possibly also in conjunction with
next-next-nearest-neighbor (NNNN) bonds of strength $J_{3}$.  The
resulting spin-$\frac{1}{2}$ $J_{1}$--$J_{2}$--$J_{3}$ model on the
honeycomb lattice, or special cases of it (e.g., when $J_{3}=0$, or
$J_{3}=J_{2}$), have been intensively investigated by many authors (see,
e.g., Refs.\ \cite{Rastelli:1979_honey,Mattsson:1994_honey,Fouet:2001_honey,Mulder:2010_honey,Wang:2010_honey,Cabra:2011_honey,Ganesh:2011_honey,Clark:2011_honey,DJJF:2011_honeycomb,Reuther:2011_honey,Albuquerque:2011_honey,Mosadeq:2011_honey,Oitmaa:2011_honey,Mezzacapo:2012_honey,PHYLi:2012_honeycomb_J1neg,Bishop:2012_honeyJ1-J2,Bishop:2012_honey_circle-phase,Li:2012_honey_full,RFB:2013_hcomb_SDVBC,Ganesh:2013_honey_J1J2mod-XXX,Zhu:2013_honey_XY,Zhang:2013_honey,Gong:2013_J1J2mod-XXX,Yu:2014_honey_J1J2mod} and references cited therein).  In particular, the
CCM has been used extensively to study the GS phase structure of the
model \cite{DJJF:2011_honeycomb,PHYLi:2012_honeycomb_J1neg,Bishop:2012_honeyJ1-J2,Bishop:2012_honey_circle-phase,Li:2012_honey_full,RFB:2013_hcomb_SDVBC}.  In these earlier studies the system was mainly
investigated via accurate calculations of the ground-state energy per
spin $E/N$, the staggered magnetization $M$, and the coefficients of susceptibility against the formation of various forms of valence-bond
crystalline order.  In the current paper we wish to extend the work to
calculate both the spin gap and the complete set of fundamental
parameters that determines the low-energy physics of this magnetic
system.

The low-energy properties of any strongly correlated system with a
spontaneous symmetry breaking are governed by the properties and
dynamics of the corresponding emergent massless Goldstone bosons.  For
such 2D HAFs as are studied here, these are simply the spin-wave (or
magnon) excitations.  The interactions between such massless Goldstone
modes, the existence of which in this case is due to the spontaneous
breaking of the SU(2) spin symmetry down to its U(1) subgroup, are
strongly constrained by symmetry considerations.  A consistent
description of the physics of the ensuing low-energy magnons in terms
of an effective theory was pioneered by Chakravarty {\it et al}.
\cite{Chakravarty:1989_magnon-chiral-PT}.

After the advent of the chiral perturbation theory ($\chi$PT) for the
(pseudo-)Goldstone pions in quantum chromodynamics, a systematic
low-energy effective field theory for magnons was quickly developed in
complete analogy
\cite{Neuberger:1989_magnon-chiral-PT,Fisher:1989_magnon-chiral-PT,Hasenfratz:1990_magnon-chiral-PT,Hasenfratz:1991_magnon-chiral-PT,Hasenfratz:1993_magnon-chiral-PT}.
The results obtained by $\chi$PT are exact, order by order, in a
consistent and systematic low-energy expansion.  They are universally
applicable to models in the same underlying symmetry class, and where
the symmetry is broken in the same way.  The corresponding low-energy
properties of such classes of systems are thus determined in terms of
a (small) set of low-energy physical parameters that enter the
effective Lagrangian or effective action, for example.  These
low-energy parameters are not themselves determined by the generic
effective field theory, but depend instead on the specific model.

\begin{figure*}[!t]
\mbox{
\subfigure[]{\scalebox{0.4}{\includegraphics{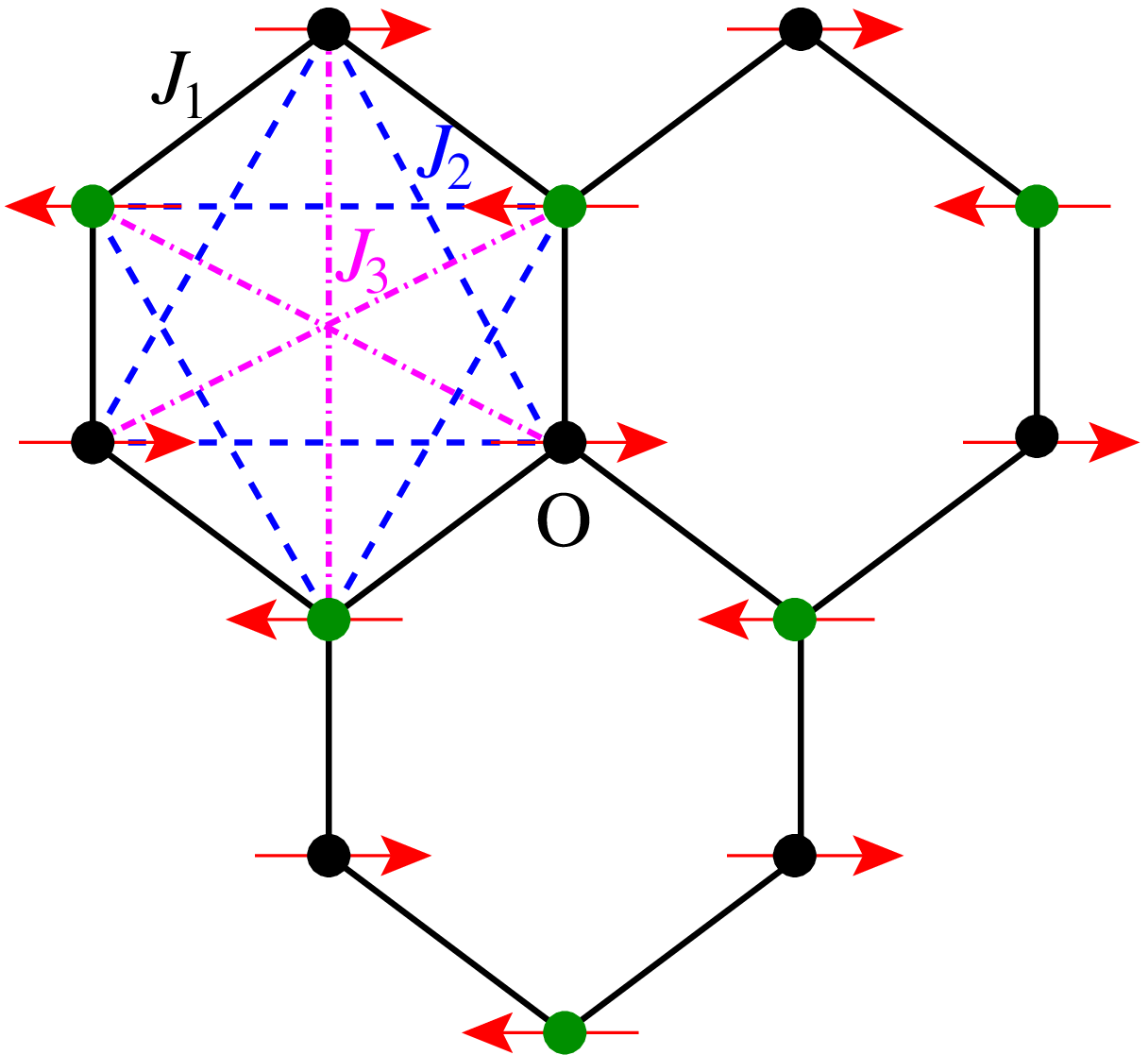}}}
\quad \subfigure[]{\scalebox{0.4}{\includegraphics{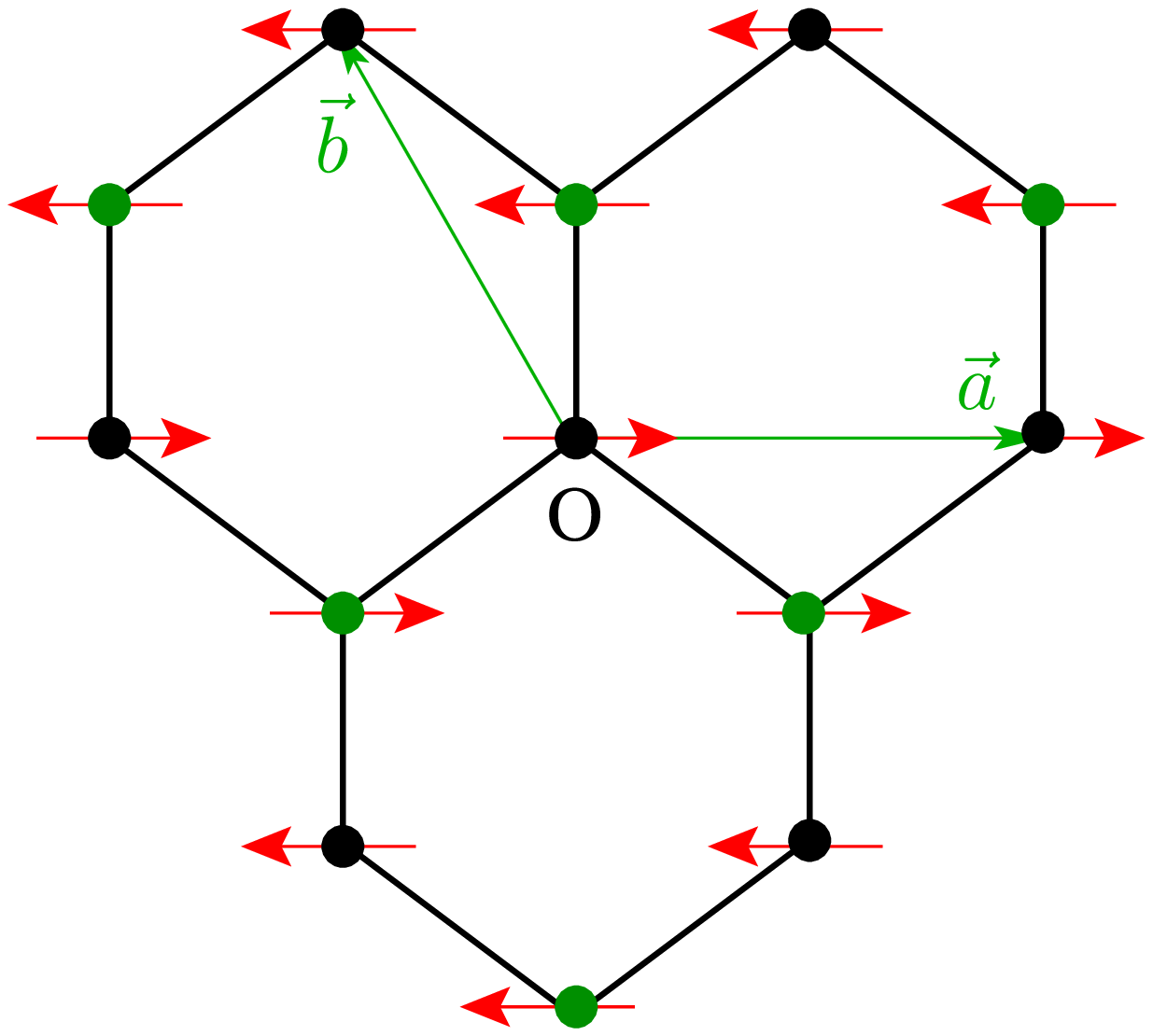}}}
\raisebox{2.28cm}{
\quad \subfigure{\scalebox{0.4}{\includegraphics{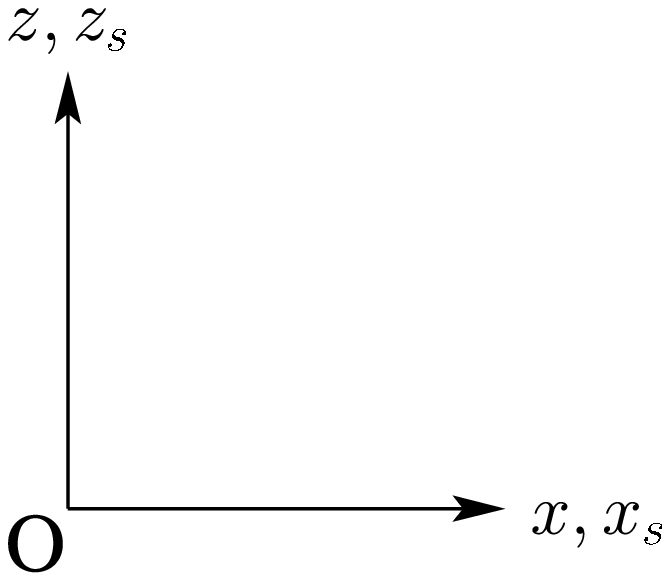}}}
}
}
\caption{(Color online) The $J_{1}$--$J_{2}$--$J_{3}$ honeycomb model
  with $J_{1}>0; J_{2}>0; J_{3}>0$, showing (a) the bonds ($J_{1} =$ 
  -----~; $J_{2} = - - -$~; $J_{3} = - \cdot -$) and the
  N\'{e}el state, and (b) triangular Bravais lattice vectors
  $\mathbf{a}$ and $\mathbf{b}$ and one of the three equivalent
  striped states.  Sites on sublattices ${\cal A}$ and ${\cal B}$ are
  shown by filled black and green circles respectively, and spins on
  the lattice are represented by the (red) arrows.}
\label{model_pattern}
\end{figure*}

The fundamental low-energy parameter set that completely determines in
this way the low-energy physics of magnetic systems comprises the GS
energy per particle $E/N$, the average local on-site magnetization
(viz., here the staggered magnetization) $M$ that plays the role of
the order parameter, the zero-field (uniform, transverse) magnetic susceptibility
$\chi$, the spin stiffness $\rho_{s}$, and the spin-wave velocity $c$.
In previous studies
\cite{DJJF:2011_honeycomb,PHYLi:2012_honeycomb_J1neg,Bishop:2012_honey_circle-phase}
of the spin-$\frac{1}{2}$ $J_{1}$--$J_{2}$--$J_{3}$ model on the
honeycomb lattice along the particularly interesting line
$J_{3}=J_{2}$, we have used the CCM implemented to high orders to give
very accurate calculations for the quantities $E/N$ and $M$ in the
magnetically ordered phases, and used them to investigate in detail
the $T=0$ phase diagram of the model.

One of our aims here is to use the CCM now to directly calculate also
the remaining low-energy parameters $\chi$ and $\rho_{s}$ (and hence
also $\hbar c = \sqrt{\rho_{s}/\chi}$, with standard AFM
hydrodynamics, and in units for $\chi$ where the gyromagnetic ratio
$g\mu_{B}/\hbar = 1$).  Together with our earlier work, a knowledge
of these quantities will provide a complete and consistent description
of the model via low-energy $\chi$PT.  Furthermore, as we shall see,
the parameters $\chi$ and $\rho_{s}$ themselves provide further
information on the $T=0$ phase structure of the model, both for the
quantum critical points (QCPs) at which quasiclassical ordering melts,
and also for the paramagnetic phases with no magnetic order into which
the system passes.  Finally, we also calculate the spin gap for the
model, using the same CCM methodology, in an attempt to provide even
more information about the model.

The plan of the remainder of the paper is as follows.  In Sec.\
\ref{model_sec} we first describe the model itself, including its
classical ($s \rightarrow \infty$) counterpart.  The CCM technique is
then briefly reviewed in Sec.\ \ref{ccm_sec}, where we concentrate on
its key features, before presenting our detailed numerical results in
Sec.\ \ref{results_sec}.  We end in Sec.\ \ref{summary_sec} with
a summary and conclusions.

\section{THE MODEL}
\label{model_sec}
The $J_{1}$--$J_{2}$--$J_{3}$ model on the honeycomb lattice is
described by the Hamiltonian
\begin{equation}
H = J_{1}\sum_{\langle i,j \rangle} \mathbf{s}_{i}\cdot\mathbf{s}_{j} + J_{2}\sum_{\langle\langle i,k \rangle\rangle} \mathbf{s}_{i}\cdot\mathbf{s}_{k} + J_{3}\sum_{\langle\langle\langle i,l \rangle\rangle\rangle} \mathbf{s}_{i}\cdot\mathbf{s}_{l}\,,
\label{eq_H}
\end{equation}
in which the index $i$ runs over all $N$ lattice sites, and the
indices $j$, $k$, and $l$ run respectively over all NN, NNN, and NNNN sites to
$i$, and where in each sum each bond is counted once and once only.
Each site $i$ of the lattice carries a spin-$\frac{1}{2}$ particle
described by the spin operator ${\bf s}_{i} \equiv (s^{x}_{i},
s^{y}_{i}, s^{z}_{i})$.  The lattice and the Heisenberg exchange bonds
are illustrated in Fig.~\ref{model_pattern}.
In the present paper we will be interested in the case where each of
the three types of bonds is AFM in nature, i.e., $J_{n} > 0$;
$n=1,2,3$.

The honeycomb lattice is bipartite with two triangular Bravais
sublattices ${\cal A}$ and ${\cal B}$.  If the lattice spacing on the
honeycomb lattice (i.e., the distance between NN sites) is $d$, then sites on sublattice ${\cal A}$ are at
positions $\mathbf{R}_{i} =
m\mathbf{a}+n\mathbf{b}=\sqrt{3}(m-\frac{1}{2}n)d
\hat{x}+\frac{3}{2}nd \hat{z}$, in terms of real-space basis vectors
$\mathbf{a}=\sqrt{3}d\hat{x}$ and
$\mathbf{b}=\frac{1}{2}d(-\sqrt{3}\hat{x}+3\hat{z})$ for the honeycomb
lattice, which is defined to lie in the $xz$ plane, as shown in Fig.\
\ref{model_pattern}.  Each unit cell $i$ at position vector
$\mathbf{R}_{i}$ comprises a pair of sites, situated at
$\mathbf{R}_{i} \in {\cal A}$ and $(\mathbf{R}_{i} + d\hat{z}) \in
{\cal B}$.  The parallelogram formed by $\mathbf{a}$ and $\mathbf{b}$
defines the honeycomb Wigner-Seitz unit cell.  The Wigner-Seitz cell
may itself equivalently be taken as being centered on a point of
sixfold symmetry so that it is bounded by the sides of a primitive
hexagon of side $d$.  The first Brillouin zone is then itself also a
hexagon, which is rotated by 90$^{\circ}$ with respect to the
Wigner-Seitz hexagon, and which has a side of length
$4\pi/(3\sqrt{3}d)$.

The classical version of the honeycomb model of Eq.\ (\ref{eq_H})
(i.e., when $s \rightarrow \infty$) has been studied by several
authors \cite{Rastelli:1979_honey,Fouet:2001_honey,Mulder:2010_honey}.
For example, Rastelli {\it et al}. \cite{Rastelli:1979_honey}
searched for coplanar or uniformly canted spin configurations that
minimize the classical energy and found the former to be energetically
favored.  Generally they correspond to spiral configurations described
by a wave vector {\bf Q}, plus an angle $\theta$ that describes the
relative orientation of the two spins in the same unit cell $i$, both
of which are now characterized by the same triangular Bravais lattice
vector $\mathbf{R}_{i}$.  The classical spins (of length $s$) in unit
cell $i$ are thus given by
\begin{equation}
\mathbf{s}_{i,\tau}=s[\cos(\mathbf{Q}\cdot\mathbf{R}_{i}+\theta_{\tau})\hat{x}_{s}+\sin(\mathbf{Q}\cdot\mathbf{R}_{i}+\theta_{\tau})\hat{z}_{s}]\,,  \label{eq_classical-spins}
\end{equation}
where $\hat{x}_{s}$ and $\hat{z}_{s}$ are two orthogonal unit vectors
that define the plane of the spins in spin space, as shown in Fig.\
\ref{model_pattern}.  The index $\tau$ labels the two sites in the
unit cell.  Clearly we may choose the angles $\theta_{\tau}$ so that
$\theta_{A}=0$ for spins on sublattice ${\cal A}$ and $\theta_{B}=\theta$,
say, for spins on sublattice ${\cal B}$.

In our regime of interest (i.e., when $J_{n}>0$; $n=1,2,3$) the
classical model has a $(T=0)$ GS phase diagram comprising three
distinct phases \cite{Fouet:2001_honey,Mulder:2010_honey}.  One value
of the spiral wave vector that minimizes the classical GS energy is
\begin{equation}
\mathbf{Q}=\frac{2}{\sqrt{3}d}\cos^{-1}\left[\frac{J_{1}-2J_{2}}{4(J_{2}-J_{3})}\right]\hat{x}\,,  \label{eq_spiral-wave-factor}
\end{equation}
together with $\theta=\pi$.  Clearly, the wave vector $\mathbf{Q}$ of
Eq.\ (\ref{eq_spiral-wave-factor}) is only properly defined when
\begin{equation}
-1 \leq \frac{J_{1}-2J_{2}}{4(J_{2}-J_{3})} \leq 1 \,,
\end{equation}
or, equivalently, when
\begin{equation}
y \leq \frac{3}{2}x-\frac{1}{4}\,; \quad y \leq \frac{1}{2}x+\frac{1}{4}\,,  \label{eq_xy}
\end{equation}
where $x \equiv J_{2}/J_{1}$ and $y \equiv J_{3}/J_{1}$.  The region
in the positive quadrant (i.e., $x \geq 0$, $y \geq 0$) of the $xy$
plane that satisfies both inequalities of Eq.\ (\ref{eq_xy}) is where
the classical $J_{1}$--$J_{2}$--$J_{3}$ model on the honeycomb lattice
has the spiral phase described by the wave vector $\mathbf{Q}$ of Eq.\
(\ref{eq_spiral-wave-factor}), and $\theta = \pi$, as the stable GS
phase.  Along the line $y = \frac{3}{2}x - \frac{1}{4}$, $\mathbf{Q}=
\mathbf{\Gamma} \equiv (0,0)$, and the phase described by Eq.\
(\ref{eq_classical-spins}) simply becomes (continuously) the collinear
N\'{e}el AFM phase illustrated in Fig.\ \ref{model_pattern}(a).  The
phase transition between the N\'{e}el and spiral phases is thus a
continuous one.  Similarly, along the line
$y=\frac{1}{2}x+\frac{1}{4}$, $\mathbf{Q}=2\pi/(\sqrt{3}d)\hat{x}$,
and the phase described by Eq.\ (\ref{eq_classical-spins}) becomes
(continuously) the collinear striped AFM phase illustrated in Fig.\
\ref{model_pattern}(b).  The phase transition between the striped and
spiral phases is thus also a continuous one.  The above two phase
boundaries meet at the point $(x,y)=(\frac{1}{2},\frac{1}{2})$ which is
the classical tricritical point.  There is a first-order phase
transition between the two collinear states (i.e., the N\'{e}el and
striped states) along the boundary line $x=\frac{1}{2}$, for
$y>\frac{1}{2}$.  In summary, when $J_{1}>0$ and $x>0$, $y>0$, the classical
$J_{1}$--$J_{2}$--$J_{3}$ model on the honeycomb lattice has three
stable GS phases (at $T=0$): a N\'{e}el AFM phase for
$y>\frac{3}{2}x-\frac{1}{4}$, $\frac{1}{6}<x<\frac{1}{2}$ and $y>0$,
$0<x<\frac{1}{6}$; a striped AFM phase for
$y>\frac{1}{2}x+\frac{1}{4}$, $x>\frac{1}{2}$; and a spiral phase for
$0<y<\frac{3}{2}x-\frac{1}{4}$, $\frac{1}{6}<x<\frac{1}{2}$ and
$0<y<\frac{1}{2}x+\frac{1}{4}$, $x>\frac{1}{2}$.

We note that the spiral and the striped states described by Eq.\
(\ref{eq_spiral-wave-factor}) and $\theta=\pi$ have two other similar
states rotated by $\pm\frac{2\pi}{3}$ in the honeycomb plane.  When $x
\rightarrow \infty$ for a fixed finite value of $y$ (i.e., when the
$J_{2}$ bond dominates), the spiral pitch angle
$\phi=\cos^{-1}[\frac{1}{4}(J_{1}-2J_{2})/(J_{2}-J_{3})] \rightarrow
\frac{2}{3}\pi$.  In this limiting case the classical model thus
becomes two HAFs on disconnected interpenetrating triangular lattices
with the classical 3-sublattice N\'{e}el ordering of NN spins oriented
at angle $\frac{2}{3}\pi$ to each other on each sublattice.  In this
limiting case the wave vector $\mathbf{Q}$ of Eq.\
(\ref{eq_spiral-wave-factor}) becomes one of the six corners of the
first Brillouin zone, $\mathbf{K}^{(1)}=4\pi/(3\sqrt{3}d)\hat{x}$.
Clearly, there are only two distinct such corner vectors, and these
describe the two inequivalent 3-sublattice N\'{e}el orderings for a
classical triangular HAF.

We also note that when the spiral pitch angle $\phi$ takes a value in
the range $\frac{2}{3}\pi < \phi \leq \pi$ the wave vector
$\mathbf{Q}$ of Eq.\ (\ref{eq_spiral-wave-factor}) lies outside the
first Brillouin zone.  It can equivalently be mapped back into the
first Brillouin zone in this case, when $\mathbf{Q}$ moves
continuously from a corner
$\mathbf{K}^{(3)}=-2\pi/(3\sqrt{3}d)\hat{x}+2\pi/(3d)\hat{z}$ of the
Brillouin zone along one of its edges to the midpoint
$\mathbf{M}^{(2)}=2\pi/(3d)\hat{z}$.  The striped state shown in Fig.\
\ref{model_pattern}(b) may thus be equivalently described by the
ordering wave vector $\mathbf{Q}=\mathbf{M}^{(2)}$ (with the relative
angle between sublattices ${\cal A}$ and ${\cal B}$ given by
$\theta=\pi$).  The other two striped states are thus given by the
wave vectors of the two other inequivalent midpoints of edges of the
first Brillouin zone,
$\mathbf{M}^{(1)}=\pi/(\sqrt{3}d)\hat{x}+\pi/(3d)\hat{z}$ and
$\mathbf{M}^{(3)}=-\pi/(\sqrt{3}d)\hat{x}+\pi/(3d)\hat{z}$ (and in both of
these cases with $\theta=0$).

Although the spin ordering of Eq.\ (\ref{eq_classical-spins}) usually
suffices to find all classical GS configurations
\cite{Villain:1977_ordByDisord}, there is an assumption that the GS
order is either unique (up to the trivial degeneracy associated with a
global spin rotation) or exhibits, at most, a discrete degeneracy.
However, exceptions can arise for special sets $\{\mathbf{Q}\}$ of
wave vectors \cite{Fouet:2001_honey,Villain:1977_ordByDisord}.  These
include the case when $\mathbf{Q}$ is equal to one half or one quarter
of a reciprocal lattice vector $\mathbf{G}$.  This is precisely the
case for the striped states where the wave vectors
$\mathbf{Q}=\mathbf{M}^{(i)}$, $i=1,2,3$ are just one half of
corresponding reciprocal lattice vectors.  As explained by Fouet {\it
  et al}. \cite{Fouet:2001_honey} the GS ordering in this case spans a
2D manifold of non-planar GS configurations, all degenerate in energy.

It is well known that classical models that exhibit such an infinitely
degenerate family (IDF) of GS phases in some region of the $T=0$ phase
space often lead to the emergence of novel quantum phases in the
corresponding quantum-mechanical model.  A typical scenario then finds
that quantum fluctuations lift this (accidental) GS degeneracy, either
completely or in part by the well-known {\it order by disorder}
mechanism
\cite{Villain:1977_ordByDisord,Villain:1980_ordByDisord,Shender:1982_ordByDisord},
so that just one or several members of the classical IDF are favored.
Indeed, thermal or quantum fluctuations do select the collinear striped
states out of the 2D IDF manifold, according to Ref.\
\cite{Fouet:2001_honey}, wherein it is also explicitly shown in an ED
study of the finite-lattice spectra that the degeneracy is lifted in favor of
the collinear ordering.

In the extreme $s=\frac{1}{2}$ quantum limit one may also expect that
quantum fluctuations might be strong enough to destroy any
quasiclassical magnetic long-range order (LRO) completely in some
region of the $T=0$ GS phase space.  The goal of finding any such
novel quantum phases with no magnetic LRO, and delimiting their region
of stability in the $T=0$ phase space, has provided the impetus for
much recent work
\cite{Rastelli:1979_honey,Mattsson:1994_honey,Fouet:2001_honey,Mulder:2010_honey,Wang:2010_honey,Cabra:2011_honey,Ganesh:2011_honey,Clark:2011_honey,DJJF:2011_honeycomb,Reuther:2011_honey,Albuquerque:2011_honey,Mosadeq:2011_honey,Oitmaa:2011_honey,Mezzacapo:2012_honey,PHYLi:2012_honeycomb_J1neg,Bishop:2012_honeyJ1-J2,Bishop:2012_honey_circle-phase,Li:2012_honey_full,RFB:2013_hcomb_SDVBC,Ganesh:2013_honey_J1J2mod-XXX,Zhu:2013_honey_XY,Zhang:2013_honey,Gong:2013_J1J2mod-XXX,Yu:2014_honey_J1J2mod}
on the spin-$\frac{1}{2}$ $J_{1}$--$J_{2}$--$J_{3}$ model on the
honeycomb lattice.  A particularly challenging, yet potentially
fruitful, regime is to consider the case $J_{3}=J_{2}\equiv \kappa
J_{1}$ since it includes the point of maximum classical frustration,
viz., the tricritical point at $\kappa = \kappa_{{\rm cl}} \equiv
\frac{1}{2}$.  Henceforth, therefore, we restrict ourselves to this
regime.

In our earlier paper \cite{DJJF:2011_honeycomb} we presented results
for the GS energy and magnetic order parameter of the
spin-$\frac{1}{2}$ $J_{1}$--$J_{2}$--$J_{3}$ model on the honeycomb
lattice along the line $J_{3}=J_{2}\equiv \kappa J_{1}$, with
$J_{1}>0$ and $\kappa > 0$, using the CCM implemented in high orders.
We found that the first-order transition in the classical ($s
\rightarrow \infty$) model at $\kappa = \kappa_{{\rm cl}} \equiv
\frac{1}{2}$ between the N\'{e}el and collinear striped AFM phases is
split into two transitions for the $s=\frac{1}{2}$ model.  The
N\'{e}el phase was found to survive for $\kappa < \kappa_{c_{1}}
\approx 0.47$, and the striped phase for $\kappa > \kappa_{c_{2}}
\approx 0.60$.  In the region $\kappa_{c_{1}} < \kappa <
\kappa_{c_{2}}$ between the two quasiclassical phases we found a
paramagnetic phase with no discernible magnetic order.  By further
calculating within the same CCM methodology, the susceptibilities of
the two AFM phases against the formation of a state with plaquette
valence-bond crystalline (PVBC) order, we concluded that the
intermediate state was one with PVBC order.  The evidence from those
calculations was that the quantum phase transition (QPT) at
$\kappa_{c_{2}}$ appears to be a first-order one, while that at
$\kappa_{c_{1}}$ is of continuous type.  Since the N\'{e}el and PVBC
phases break different symmetries, we concluded that the quantum
transition at $\kappa_{c_{1}}$ between these two phases is of the
deconfined type
\cite{Senthil:2004_Science_deconfinedQC,*Senthil:2004_PRB_deconfinedQC}.

Our aim in the present paper is to shed further light on the model by
calculating other physical properties within the same CCM methodology
as used previously.  Firstly, in order to gain more evidence about the
nature of the intermediate phase we calculate the spin gap, i.e., the
energy gap between the ground state and the lowest-lying (magnon)
triplet excitation.  Secondly, as discussed previously, we also
calculate for both quasiclassical phases the spin stiffness,
$\rho_{s}$, and the zero-field (uniform, transverse) magnetic susceptibility, $\chi$,
in order both to provide a complete set of low-energy parameters for
both phases of the model and to use these parameters to provide more
numerical evidence for the two QPTs at $\kappa_{c_{1}}$ and
$\kappa_{c_{2}}$.

\begin{figure}[!tb]
\mbox{
\subfigure[]{\scalebox{0.3}{\includegraphics{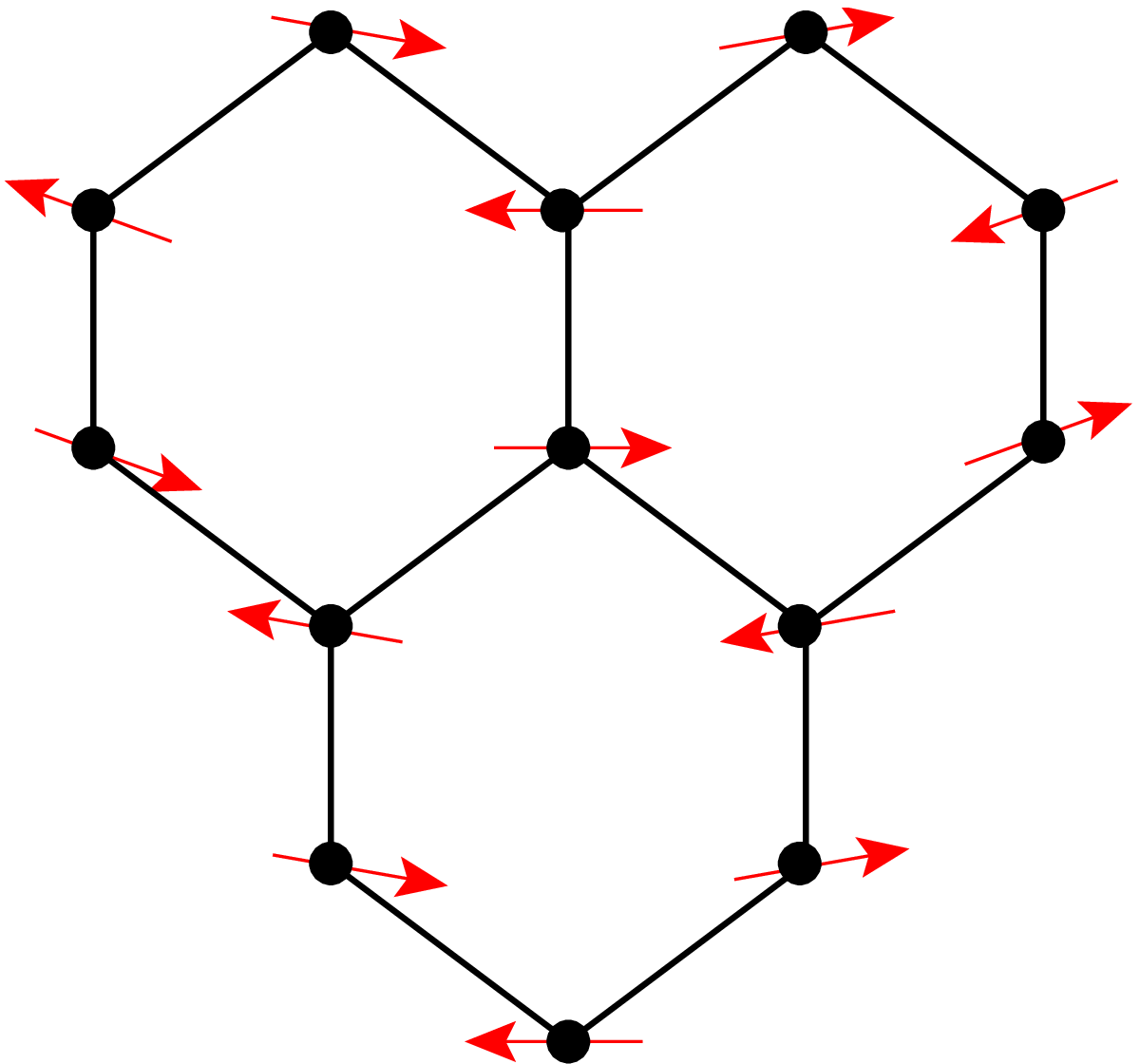}}}
\quad \subfigure[]{\scalebox{0.3}{\includegraphics{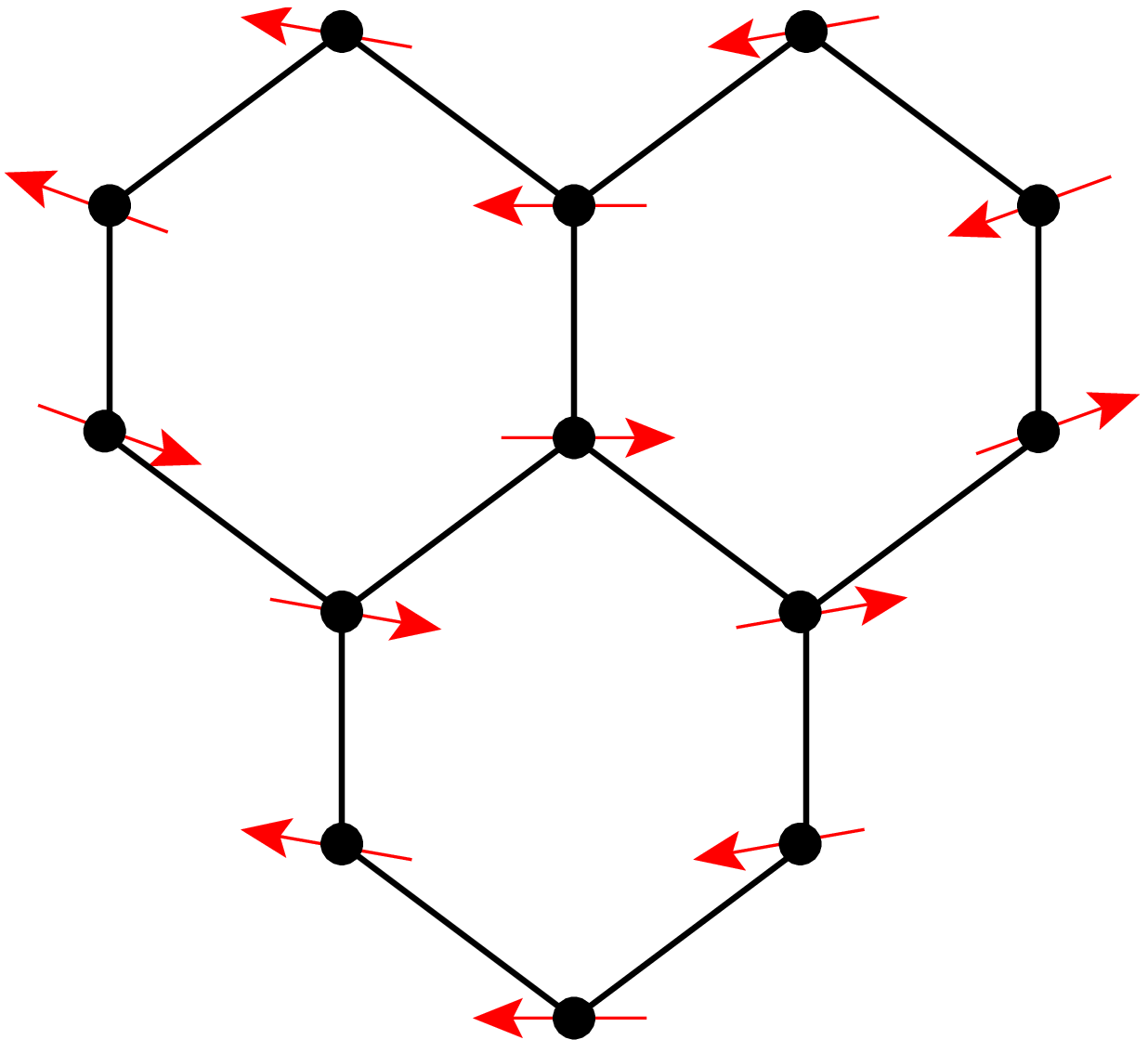}}}
}
\caption{(Color online) The twisted reference states for the calculation of the spin stiffness coefficient, $\rho_{s}$, for the $J_{1}$--$J_{2}$--$J_{3}$ honeycomb model, showing the twist applied in the $x$ direction to 
  the (a) N\'{e}el and (b) striped states.
  The spins on lattice sites \textbullet \hspace{0.01cm} are represented by the (red) arrows.}
\label{pattern_sStiff}
\end{figure}

The spin stiffness (or helicity modulus) of a spin-lattice system is a
measure of the energy required to rotate the order parameter of a
magnetically ordered thermodynamic system by an (infinitesimal) angle
$\theta$ per unit length in a given direction.  Thus, if $E(\theta)$
is the GS energy as a function of the imposed twist and $N$ is the
number of lattice sites, we have
\begin{equation}
\frac{E(\theta)}{N}=\frac{E(\theta=0)}{N} + \frac{1}{2}\rho_{s}\theta^{2} + O(\theta^{4})\,.  \label{eq_GS-E_theta}
\end{equation}
Note that $\theta$ has the dimensions of an inverse length.  In the
thermodynamic limit ($N \rightarrow \infty$) a nonzero (positive)
value of $\rho_{s}$ implies that the system has magnetic long-range
order (LRO), while the magnetic LRO melts at the point where $\rho_{s}
\rightarrow 0$.  Clearly, for the N\'{e}el AFM state of Fig.\
\ref{model_pattern}(a), for which the ordering wave vector $\mathbf{Q}
= \mathbf{\Gamma} = (0,0)$, the quantity $\rho_{s}$ is independent of
the direction of the applied twist, whereas for the particular striped
AFM state shown in Fig.\ \ref{model_pattern}(b), for which
$\mathbf{Q}=2\pi/(\sqrt{3}d)\hat{x}$, the physically relevant
direction in which to apply the twist is the $x$ direction.  Figure
\ref{pattern_sStiff} thus illustrates the two twisted AFM states
(i.e., the N\'{e}el and striped states) used in our calculations for
the spin stiffness coefficient.
The definition of Eq.\ (\ref{eq_GS-E_theta}) easily leads to the
corresponding values of $\rho_{s}$ for the classical $(s \rightarrow
\infty)$ $J_{1}$--$J_{2}$--$J_{3}$ model on the honeycomb lattice,
\begin{equation}
\rho^{{\rm N\acute{e}el}}_{s;\,{\rm cl}}=\frac{3}{4}(J_{1}-6J_{2}+4J_{3})d^{2}s^{2}\,, \label{sStiff_neel_classical}
\end{equation}
and
\begin{equation}
\rho^{{\rm striped}}_{s;\,{\rm cl}}=\frac{3}{4}(-J_{1}-2J_{2}+4J_{3})d^{2}s^{2}\,. \label{sStiff_stripe_classical}
\end{equation}
The lines along which the spin stiffness coefficients given by Eqs.\
(\ref{sStiff_neel_classical}) and (\ref{sStiff_stripe_classical})
vanish are, as expected, just the classical phase boundaries for the
two AFM states with the spiral state.

It is also interesting to note that in the limiting case $J_{3}
\rightarrow \infty$, for fixed values of $J_{1}$ and $J_{2}$, the
$J_{1}$--$J_{2}$--$J_{3}$ model reduces to four decoupled
honeycomb-lattice HAFs, each with NN coupling $J_{3}$ and lattice
spacing 2$d$.  Thus, the GS energy in the case $\{J_{1}=0,\, J_{2}=0,\,
J_{3}=1\}$ should be equal to that of the case when $\{J_{1}=1,\, J_{2}=0,\,
J_{3}=0\}$.  Similarly, the spin stiffness in the former limit should
equal four times that in the latter limit, due to the doubling of the
lattice size of each of the four decoupled honeycomb sublattices.
Just, as this result holds for the classical model ($s \rightarrow
\infty$), from Eq.\ (\ref{sStiff_neel_classical}), so it should also
hold for general values of the spin quantum number $s$.

\begin{figure}[!tb]
\mbox{
\subfigure[]{\scalebox{0.3}{\includegraphics{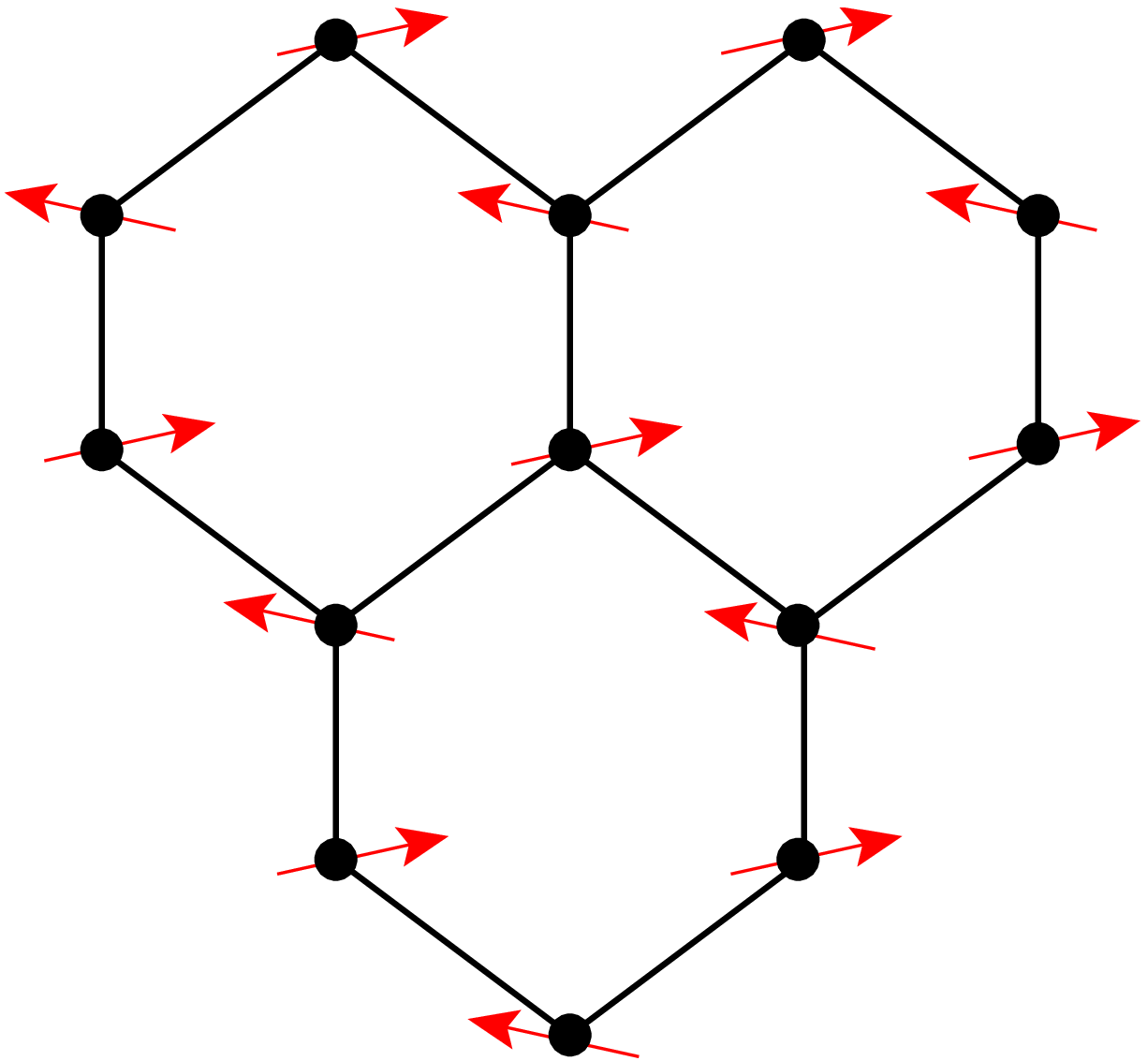}}}
\quad \subfigure[]{\scalebox{0.3}{\includegraphics{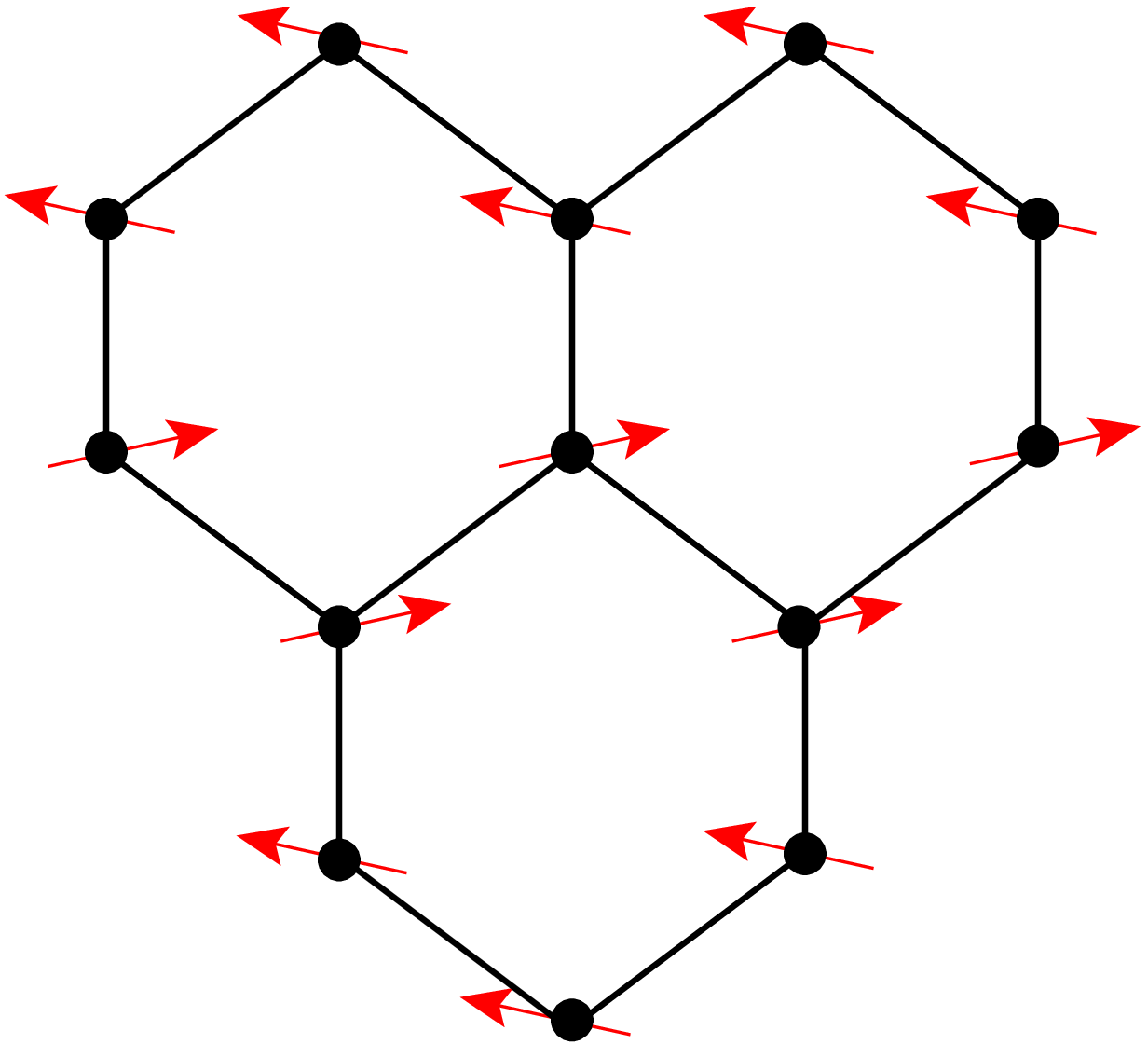}}}
}
\caption{(Color online) The canted reference states for the calculation of the zero-field magnetic susceptibility, $\chi$, for the $J_{1}$--$J_{2}$--$J_{3}$ honeycomb model.  The external magnetic field is applied in the $z_{s}$ direction to 
  the (a) N\'{e}el and (b) striped states, shown in Fig.\ \ref{model_pattern}.
  The spins on lattice sites \textbullet \hspace{0.01cm} are represented by the (red) arrows.}
\label{pattern_M_ExtField}
\end{figure}

In order to calculate the zero-field magnetic susceptibility, $\chi$,
we now place our system in an external transverse magnetic field
$\mathbf{h}$.  For the two collinear AFM states shown in Fig.\
\ref{model_pattern}, both with spins aligned along the $x_{s}$
direction, we apply the field in the $z_{s}$ direction,
$\mathbf{h}=h\hat{z_{s}}$.  The Hamiltonian $H=H(h=0)$ of Eq.\
(\ref{eq_H}) thus becomes
\begin{equation}
H(h)=H(h=0) + h\sum_{l} s^{z}_{l}\,,  \label{eq_H-h}
\end{equation}
in units where the gyromagnetic ratio $g\mu_{B}/\hbar=1$.  In the
presence of the field, the spins will now cant at an angle $\alpha$ to
the $x_{s}$ axis with respect to their zero-field configurations, as
shown in Fig.\ \ref{pattern_M_ExtField} for the two classical AFM states
illustrated in Fig.\ref{model_pattern}.  
The classical ($s \rightarrow
\infty$) value of $\alpha$ for each of the states is easily calculated
by minimizing the classical energy, $E = E(h)$, corresponding to Eq.\
(\ref{eq_H-h}) with respect to $\alpha$.  The uniform magnetic
susceptibility is then defined, as usual, by
\begin{equation}
\chi({h})=-\frac{1}{N}\frac{{\rm d}^{2}E}{{\rm d}h^{2}}\,,
\end{equation}
and its zero-field limit as $\chi \equiv \chi(0)$.  The corresponding
analog of Eq.\ (\ref{eq_GS-E_theta}) is thus,
\begin{equation}
\frac{E(h)}{N}=\frac{E(h=0)}{N}-\frac{1}{2}\chi h^{2} + O(h^{4})\,.
\end{equation}
These relations readily yield the values of $\chi$ for the two collinear AFM states of the classical ($s \rightarrow \infty$) $J_{1}$--$J_{2}$--$J_{3}$ model on the honeycomb lattice,
\begin{equation}
\chi^{{\rm N\acute{e}el}}_{{\rm cl}}=\frac{1}{6(J_{1}+J_{3})}\,, \label{chi_neel_classical}
\end{equation}
and
\begin{equation}
\chi^{{\rm striped}}_{{\rm cl}}=\frac{1}{2(J_{1}+4J_{2}+3J_{3})}\,,  \label{chi_stripe_classical}
\end{equation}
with both parameters independent of $s$ in the classical case.  The
two values of $\chi$ become equal (but nonzero, note) along the line
$J_{2}=\frac{1}{2}J_{1}$ (independent of $J_{3}$), which is the
classical phase boundary between the two AFM states.

We note that our definitions for both $\rho_{s}$ and $\chi$ are per
unit site, as is more usual for a discrete lattice description.  On the
other hand, in continuum field-theoretic terms, such as in
descriptions using $\chi$PT, it is more natural to define
corresponding quantities, $\bar{\rho_{s}}$ and $\bar{\chi}$, per unit
area.  Since the honeycomb lattice has $4/(3\sqrt{3}d^{2})$ sites per
unit area, $\rho_{s}=\frac{3}{4}\sqrt{3}d^{2}\bar{\rho_{s}}$ and
$\chi=\frac{3}{4}\sqrt{3}d^{2}\bar{\chi}$.

\section{THE COUPLED CLUSTER METHOD}
\label{ccm_sec}
The CCM is one of only a very few universal methods of {\it ab initio}
quantum many-body theory (QMBT) that are capable of systematic
improvement in well-defined hierarchies of approximations.  It is
nowadays one of the most pervasive, most powerful, and most accurate
at attainable levels of computational implementation, of all fully
microscopic formulations of QMBT.  It has been applied with
considerable success to a wide range of physical systems
\cite{Bishop:1987_ccm,Bartlett:1989_ccm,Arponen:1991_ccm,Bishop:1991_TheorChimActa_QMBT,Bishop:1998_QMBT_coll},
ranging from the electron gas to atoms and molecules of interest in
quantum chemistry, from nuclear matter and finite atomic nuclei to
strongly interacting quantum field theories, and from models in
quantum optics, quantum electronics, and solid state optoelectronics
to diverse condensed matter systems.  The CCM has thus yielded accurate
numerical results for a very wide range of both finite and extended
systems defined either in a spatial continuum or on a regular discrete
lattice.  Of particular relevance to the present paper is the fact
that the method has already been applied with demonstrable success to
a large number of spin-lattice problems of interest in quantum
magnetism (and see, e.g., Refs.\
\cite{Bishop:1998_honey,DJJFarnell:2014_archimedeanLatt,DJJF:2011_honeycomb,PHYLi:2012_honeycomb_J1neg,Bishop:2012_honeyJ1-J2,Bishop:2012_honey_circle-phase,Li:2012_honey_full,RFB:2013_hcomb_SDVBC,RoHe:1990_1D-2D,Bishop:1992_JPCM4_LSUBn,Zeng:1998_SqLatt_TrianLatt,Fa:2004_QM-coll,SEKruger:2006_spinStiff,Darradi:2008_J1J2mod,Farnell:2009_Xcpty_ExtMagField,Gotze:2011_kagome,Bishop:2014_honey_XY,Li:2015_j1j2-triang,Richter:2015_ccm_J1J2sq_spinGap}
and references cited therein).  

Two characteristic and rather unique features of the CCM that are
immediately worthy of mention are: (i) its ability to deal with
infinite ($N \rightarrow \infty$) systems from the outset, hence
obviating the need for any finite-size scaling, which is required in
most competing methods; and (ii) the fact that it exactly preserves
the important Hellmann-Feynman theorem at {\it all} levels of
approximate implementation.  The method is implemented in practice, as
we explain below, at various levels of approximation in a well-defined
truncation hierarchy, with each level specified by a truncation index
$n=1,2,3,\cdots$.  The {\it only} approximation made is then to
extrapolate the values obtained for the physical observables at the
$n$th level to the $n \rightarrow \infty$ limit where the CCM becomes
exact.

The CCM has been described in detail in earlier papers (and see, e.g.,
Refs.\
\cite{Arponen:1991_ccm,Bishop:1998_QMBT_coll,Bishop:1991_TheorChimActa_QMBT,Bishop:1992_JPCM4_LSUBn,Zeng:1998_SqLatt_TrianLatt,DJJFarnell:2014_archimedeanLatt,Bishop:2014_honey_XY,Li:2015_j1j2-triang}
and references cited therein), and hence we briefly outline only its
most salient features here.  Every implementation of the method begins
with the choice of a suitable normalized reference (or model) state,
with respect to which the quantum correlations present in the
exact GS phase under study can then be incorporated at the next stage.
For this study suitable choices of the model state $|\Phi\rangle$ will
be the two quasiclassical AFM states (viz., the N\'{e}el and collinear
striped states) shown in Figs.\ \ref{model_pattern}(a) and
\ref{model_pattern}(b).

The exact GS ket- and bra-state wave functions, which satisfy the
respective Schr\"{o}dinger equations,
\begin{equation}
H|\Psi\rangle=E|\Psi\rangle\,; \quad \langle\tilde{\Psi}|H = E\langle\tilde{\Psi}|\,,  \label{schrodinger_eq}
\end{equation}
are chosen to have normalization conditions such that
$\langle\tilde{\Psi}|\Psi\rangle = \langle{\Phi}|\Psi\rangle =
\langle{\Phi}|\Phi\rangle = 1$.  These exact states are then
parametrized in terms of the respective model state as
\begin{equation}
|\Psi\rangle=e^{S}|\Phi\rangle\,; \quad \langle\tilde{\Psi}|=\langle\Phi|\tilde{S}e^{-S}\,,  \label{exp_para}
\end{equation}
with the exponential parametrization being a key characteristic
feature of the CCM.  The two correlation operators are then themselves
formally decomposed as
\begin{equation}
S=\sum_{I\neq 0}{\cal S}_{I}C^{+}_{I}\,; \quad \tilde{S}=1+\sum_{I\neq 0}\tilde{{\cal S}}_{I}C^{-}_{I}\,,  \label{correrlation_oper}
\end{equation}
where we define $C^{+}_{0}\equiv 1$ to be the identity operator, and
where the set index $I$ denotes a complete set of single-particle
configurations for all of the particles.  In our present spin-lattice
application it defines a specific multispin-flip configuration with
respect to the model state $|\Phi\rangle$, such that
$C^{+}_{I}|\Phi\rangle$ is the corresponding wave function for this
configuration of spins.  The model state $|\Phi\rangle$ thus acts as s
fiducial (or cyclic) vector or, in other words, as a generalized
vacuum state, with respect to the complete set of mutually commuting creation operators
$\{C^{+}_{I}\}$, and which hence satisfy the conditions
$\langle\Phi|C^{+}_{I} = 0 = C^{-}_{I}|\Phi\rangle\,, \quad \forall I
\neq 0$, where the destruction operators $C^{-}_{I} \equiv
(C^{+}_{I})^{\dagger}$.

It is very convenient for spin-lattice problems to consider each
lattice site $i$ as totally equivalent to all others, whatever the
choice of model state $|\Phi\rangle$, and one simple way to do this is
to make a passive rotation of each spin so that in its own local
spin-coordinate frame it points, say, in the downward (i.e., negative
$z_{s}$) direction as in the spin coordinate frame shown in Fig.\
\ref{model_pattern}.  Such choices of local spin coordinates clearly
do not affect the basic SU(2) spin commutation relations.  However,
all independent-spin product model states now take the universal form
$|\Phi\rangle =
|\downarrow\downarrow\downarrow\cdots\downarrow\rangle$.  Thus, $C^{+}_{I}$ can simply be expressed as a
product of single-spin raising operators, $s^{+}_{k}
\equiv s^{x}_{k}+is^{y}_{k}$, such that $C^{+}_{I} \equiv
s^{+}_{k_{1}}s^{+}_{k_{2}}\cdots s^{+}_{k_{n}};\; n=1,2,\cdots , 2sN$.  For the
present study, where we consider $s=\frac{1}{2}$, each site index
included in the corresponding set index $I \equiv
\{k_{1},k_{2},\cdots , k_{n};\; n=1,2,\cdots , 2sN\}$ may appear no
more than once.  Once the local spin coordinates have been chosen for
the particular model state $|\Phi\rangle$, the Hamiltonian $H$ simply
needs to be re-expressed in terms of them.

In principle, what remains is then to calculate the CCM correlation
coefficients $\{{\cal S}_{I}, \tilde{{\cal S}}_{I}\}$.  This is achieved
by minimization of the GS energy expectation functional,
\begin{equation}
\bar{H}=\bar{H}[{\cal S}_{I},{\tilde{\cal S}_{I}}]\equiv
\langle\Phi|\tilde{S}e^{-S}He^{S}|\Phi\rangle\,,  \label{GS_E_xpect_funct}
\end{equation}
with respect to each of the coefficients $\{{\cal S}_{I},{\tilde{\cal
    S}}_{I}\,; \forall I \neq 0\}$.  From Eqs.\
(\ref{correrlation_oper}) and (\ref{GS_E_xpect_funct}), variation with
respect to ${\tilde{\cal S}}_{I}$ yields the coupled set of non-linear
equations,
\begin{equation}
\langle\Phi|C^{-}_{I}e^{-S}He^{S}|\Phi\rangle = 0\,, \quad \forall I \neq 0\,,  \label{ket_eq}
\end{equation}
for the coefficients $\{{\cal S}_{I}\}$.  Similarly, variation of Eq.\
(\ref{GS_E_xpect_funct}) with respect to ${\cal S}_{I}$ yields the
corresponding set of linear equations
\begin{equation}
\langle\Phi|\tilde{S}e^{-S}[H,C^{+}_{I}]e^{S}|\Phi\rangle=0\,, \quad \forall I \neq 0\,,  \label{bra_eq}
\end{equation}
for the coefficients $\{{\tilde{{\cal S}}}_{I}\}$, once Eq.\ (\ref{ket_eq})
has been solved for the coefficients $\{{\cal S}_{I}\}$.  The GS energy
eigenvalue $E$, which is the value of $\bar{H}$ at the minimum, is
then simply
\begin{equation}
E=\langle\Phi|e^{-S}He^{S}|\Phi\rangle=\langle\Phi|He^{S}|\Phi\rangle\,.
\end{equation}
Equation (\ref{bra_eq}) may then be written in the equivalent form,
\begin{equation}
\langle\Phi|\tilde{S}(e^{-S}He^{S}-E)C^{+}_{I}|\Phi\rangle=0\,, \quad \forall I \neq 0\,,  \label{bra_eq_alternative}
\end{equation}
of a set of generalized linear eigenvalue equations for the
coefficients $\{\tilde{\cal S}_{I}\}$.

Having built in the correlations into our CCM parametrization of the GS wave function $|\Psi\rangle$, it now suffices to apply a linear excitation operator $X^{e}$ to $|\Psi\rangle$ to parametrize the excited-state wave function $|\Psi_{e}\rangle$ as
\begin{equation}
|\Psi_{e}\rangle = X^{e}e^{S}|\Phi\rangle\,, \quad X^{e}=\sum_{I\neq 0}{\cal X}^{e}_{I}C^{+}_{I}\,.  \label{xcited_wave_funct}
\end{equation} 
By suitably combining the GS Schr\"{o}dinger equation (\ref{schrodinger_eq}) with its excited-state counterpart,
\begin{equation}
H|\Psi_{e}\rangle=E_{e}|\Psi\rangle\,, \label{schrodinger_eq_xcited}
\end{equation}
and realizing that the operators $X^{e}$ and $S$ commute, by
construction, we readily deduce the equation,
\begin{equation}
e^{-S}[H, X^{e}]e^{S}|\Phi\rangle = \Delta_{e}X^{e}|\Phi\rangle\,,  \label{eq_xcited}
\end{equation}
where $\Delta_{e} \equiv E_{e}-E$ is the excitation energy.  By
taking the overlap of Eq.\ (\ref{eq_xcited}) with the state $\langle
\Phi|C^{-}_{I}$ we find 
\begin{equation}
\langle\Phi|C^{-}_{I}e^{-S}[H,X^{e}]e^{S}|\Phi\rangle = \Delta_{e} {{\cal X}}^{e}_{I}\,, \quad \forall I \neq 0\,,  \label{ket_eq_xcited}
\end{equation}
when the states labelled by the indices $\{I\}$ are, as usual,
orthonormalized, $\langle \Phi|C^{-}_{I}C^{+}_{J}|\Phi\rangle = \delta
(I,J)$.  The generalized eigenvalue equation (\ref{ket_eq_xcited}) is
then solved for $\Delta_{e}$.

In the present case the configurations chosen in the expansion of Eq.\
(\ref{xcited_wave_funct}) for the excitation operator $X^{e}$ are
restricted to those that change the $z$ component of the total spin,
$S^{z}_{T}$, by one.  The value we thus obtain for $\Delta_{e}$ is then
the spin-triplet gap, which we henceforth denote as $\Delta$.

Up to this point no approximations have been made.  Nevertheless, the
equations (\ref{ket_eq}) for the coefficients $\{{\cal S}_{I}\}$ are
intrinsically nonlinear and one may wonder if truncations are needed
in the evaluation of the exponential functions.  We note that these
appear, however, only in the combination of the similarity transform
$e^{-S}He^{S}$ of the Hamiltonian, which may be expanded in terms of
the well-known nested commutator sum.  Another key feature of the CCM
is that this otherwise infinite sum now actually terminates {\it
  exactly} with the double commutator term.  This is due to the basic
SU(2) commutation relations and because all of the terms in Eq.\
(\ref{correrlation_oper}) comprising $S$ commute with one another and
are simple products of spin-raising operators.  All terms in the
expansion of $e^{-S}He^{S}$ are thus linked, and the Goldstone linked
cluster theorem is exactly preserved, even if the expansion of Eq.\
(\ref{correrlation_oper}) is truncated in any way, thereby
guaranteeing size-extensivity at any such level of truncation and our
ability to work from the outset in the thermodynamic ($N \rightarrow
\infty$) limit.  Similar considerations also apply to Eqs.\
(\ref{bra_eq}) and (\ref{ket_eq_xcited}).

Thus, the only approximation made in practice for the GS calculation
is to restrict the set of indices $\{I\}$ retained in the expansions
of Eq.\ (\ref{correrlation_oper}) for the operators $\{S,\tilde{S}\}$.
Here we utilize the well-tested localized (lattice-animal-based
subsystem) LSUB$n$ scheme used in our earlier work on this model
\cite{DJJF:2011_honeycomb,PHYLi:2012_honeycomb_J1neg,Bishop:2012_honey_circle-phase}
and in many other studies too.  The LSUB$n$ scheme is defined so that
at the $n$th level of approximation we retain all multispin-flip
configurations $\{I\}$ in Eq.\ (\ref{correrlation_oper}) that are
defined over $n$ or fewer contiguous lattice sites.  Such cluster
configurations are said to be contiguous in this sense if every site
in the cluster is NN to at least one other.  The number,
$N_{f}=N_{f}(n)$, of such distinct fundamental configurations may be
reduced by utilizing the space- and point-group symmetries of the
model, together with any conservation laws that pertain to both the
Hamiltonian and the specific model state being used.  Even so, the
number $N_{f}(n)$ increases rapidly as the truncation index $n$ is
increased, and the need eventually arises to use massive
parallelization together with supercomputing resources for the
highest-order calculations \cite{Zeng:1998_SqLatt_TrianLatt,ccm_code}.

For the excited-state calculation of the spin gap $\Delta$ the choice
of clusters for the excitation operator $X^{e}$ which are retained in
Eq.\ (\ref{xcited_wave_funct}) is different from that of the GS, since
we restrict ourselves now to those that change the $z$ component,
$S^{z}$, of total spin by one unit.  Nevertheless, we use the same
LSUB$n$ scheme for both the GS and excited-state calculations, thereby
ensuring comparable accuracy for both.  The number $N_{f}=N_{f}(n)$ at
a given level $n$ of truncation is appreciably higher for the
excited-state than for the corresponding GS calculation.  Nevertheless
the corresponding CCM equations have been solved for the present
model using both quasiclassical (N\'{e}el and striped) AFM states as
model states in LSUB$n$ approximations with $n \leq 12$.

Clearly, the CCM LSUB$n$ approximations become exact, by construction,
in the $n \rightarrow \infty$ limit.  There exist well-tested,
accurate extrapolation rules for the GS quantities $E/N$ and $M$, as
we have described and used in our earlier paper
\cite{DJJF:2011_honeycomb} for this model.  Similarly, for the spin
gap, we use the extrapolation scheme \cite{Richter:2015_ccm_J1J2sq_spinGap,Kruger:2000_JJprime},
\begin{equation}
\Delta(n) = d_{0}+d_{1}n^{-1}+d_{2}n^{-2}\,,   \label{Eq_spin_gap}
\end{equation}
to obtain the extrapolated value $\Delta \equiv \Delta(\infty)=d_{0}$
from the CCM LSUB$n$ approximations, $\Delta(n)$.  Similar schemes have
also been successfully used previously for both the spin stiffness
$\rho_{s}$ \cite{SEKruger:2006_spinStiff,Darradi:2008_J1J2mod},
\begin{equation}
\rho_{s}(n) = s_{0}+s_{1}n^{-1}+s_{2}n^{-2}\,,   \label{Eq_sstiff}
\end{equation}
and the zero-field magnetic susceptibility, $\chi$
\cite{Darradi:2008_J1J2mod,Farnell:2009_Xcpty_ExtMagField},
\begin{equation}
\chi(n) = x_{0}+x_{1}n^{-1}+x_{2}n^{-2}\,.   \label{Eq_X}
\end{equation}
In the latter case, as a check on the validity and accuracy of the
scheme, we also utilize the completely unbiased scheme,
\begin{equation}
\chi(n) = \bar{x}_{0}+\bar{x}_{1}n^{-\nu}\,,  \label{Eq_X_nu}
\end{equation}
in which the leading exponent $\nu$ is itself a fitting parameter,
along with $\bar{x}_{0}$ and $\bar{x}_{1}$.  Finally, since the
lowest-order LSUB$n$ approximants (particularly that with $n=2$) are
less likely to conform well to these extrapolation rules than those
with higher values of $n$, and also since the hexagon is the
fundamental structural element of the honeycomb lattice, we prefer to
use LSUB$n$ data with $n \geq 6$, whenever practicable, to perform
each of the extrapolations in practice.

\section{RESULTS}
\label{results_sec}
In our earlier work on the spin-$\frac{1}{2}$
$J_{1}$--$J_{2}$--$J_{3}$ model on the honeycomb lattice along the
line $J_{3}=J_{2}$ $(= \kappa J_{1})$ in phase space
\cite{DJJF:2011_honeycomb} we employed the CCM and computed LSUB$n$
results for both the GS energy per spin, $E/N$, and magnetic order
parameter, $M$, with values of the truncation index $n \leq 12$, using
both the N\'{e}el and striped collinear AFM states as CCM model
states.  Although the number of fundamental configurations, $N_{f}$,
at a given LSUB$n$ level of approximation is greater for the triplet
excited state than for the ground state for the corresponding CCM
calculations based on both model states, we are still able to
calculate the spin gap $\Delta$ at LSUB$n$ levels with $n \leq 12$,
even with the increased computational difficulty.  We are thus able to
achieve comparable accuracy for both the ground and excited states.

\begin{figure*}
\mbox{
\subfigure[]{\scalebox{0.3}{\includegraphics[angle=270]{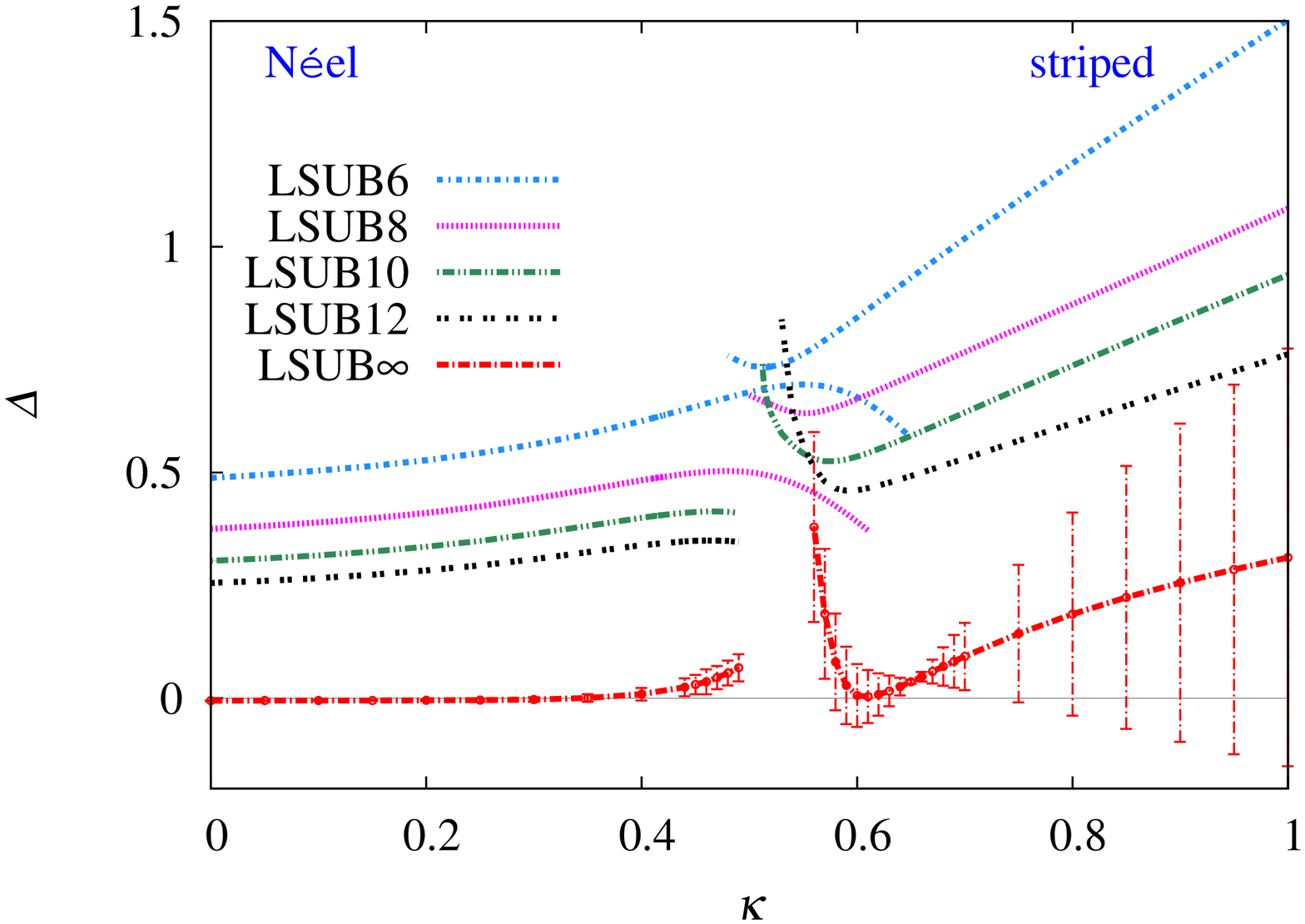}}}
\quad \subfigure[]{\scalebox{0.3}{\includegraphics[angle=270]{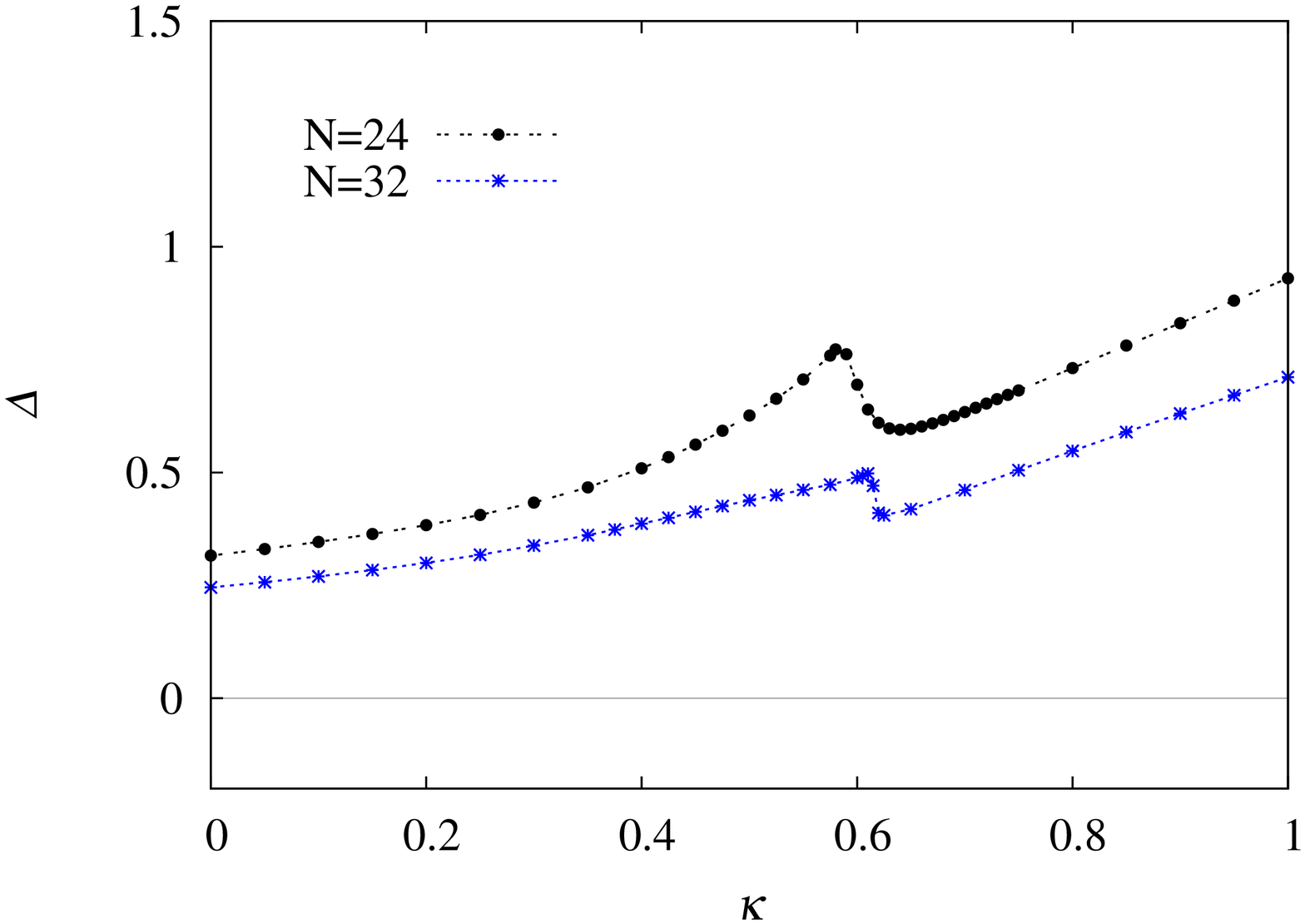}}}
}
\caption{(Color online) The spin gap $\Delta$ versus the frustration parameter $\kappa
  \equiv J_{2}/J_{1}$, for the spin-$\frac{1}{2}$ $J_{1}$--$J_{2}$--$J_{3}$
  model on the honeycomb lattice (with $J_{1}=1$, $J_{3}=J_{2}>0$).  (a) CCM results based on the N\'{e}el and the striped
  AFM states as CCM model states.  The LSUB$n$ results are shown with
  $n=\{6,8,10,12\}$.  The extrapolated LSUB$\infty$ results
  using Eq.\ (\ref{Eq_spin_gap}) are shown, together with the error bars associated with the fit.  (b) ED results for lattices with $N=24$ and $N=32$ spins.}
\label{E_gap}
\end{figure*}

We show in Fig.\ \ref{E_gap}(a) our ``raw'' LSUB$n$ results for the
spin gap $\Delta$ with $n=\{6,8,10,12\}$, based on both the N\'{e}el
and striped collinear states used separately as the CCM model state.
We also show the corresponding (LSUB$\infty$) extrapolated values, as
obtained using the scheme of Eq.\ (\ref{Eq_spin_gap}).  For the
N\'{e}el results our method of solution is as follows.  We first solve
for the pure Heisenberg AFM in the case $\kappa=0$, where we find a
stable physical solution to the CCM equations at each LSUB$n$ level of
truncation.  For a given value of the truncation index $n$ the
corresponding LSUB$n$ solution is tracked as the frustration parameter
$\kappa$ is increased incrementally, just to the point where this
stable solution terminates, as shown in Fig.\ \ref{E_gap}(a).  These
termination points for the excited-state CCM equations are completely
analogous to those also found for the GS CCM equations, which have
been well described and documented previously (see, e.g., Refs.\
\cite{Bishop:2012_honeyJ1-J2,RFB:2013_hcomb_SDVBC,Fa:2004_QM-coll,Bishop:2014_honey_XY}).
They are direct manifestations of the respective QCP in the physical
system under study, at which the corresponding form of magnetic order
in the associated model state melts.  As is usually the case, we find
that each LSUB$n$ N\'{e}el solution with a fixed (finite) value of $n$
terminates at a {\it higher} value of $\kappa$ than the corresponding
actual critical value, $\kappa_{c_{1}}$, which is just the
LSUB$\infty$ limiting value.  The outcome is that we may thereby
consider a range for the parameter $\kappa$ that is appreciably beyond
the (seemingly continuous) transition at $\kappa_{c_{1}}$ from the
N\'{e}el phase to the quantum paramagnetic phase.

On the other side of the phase diagram we similarly track each LSUB$n$
solution based on the collinear striped state as CCM model state from
high values of $\kappa$ down to some respective lower transition
point, corresponding to the actual transition at $\kappa_{c_{2}}$.
Again, for each value of the transition index $n$, we may enter into
the region below the (seemingly first-order) transition at
$\kappa_{c_{2}}$ into the quantum paramagnetic phase.  Of course, if
the transition at $\kappa_{c_{2}}$ is indeed of first-order type, as
seems likely from all prior available evidence, one might possibly query the
validity of our CCM results based on the striped collinear state in
the region $\kappa < \kappa_{c_{2}}$.

  In Fig.\ \ref{E_gap}(a) for the LSUB$\infty$ extrapolated values of
  $\Delta$ based on our LSUB$n$ results with $n=\{6,8,10,12\}$, we
  also show the error bars associated with the assumed scheme of Eq.\
  (\ref{Eq_spin_gap}).  It is clear that the fit on the N\'{e}el
  side of the phase diagram is markedly better than that on the
  striped collinear side.  Very interestingly, very similar behavior
  was also observed in a recent CCM study of the spin gap in the
  spin-$\frac{1}{2}$ $J_{1}$--$J_{2}$ model on the square lattice
  \cite{Richter:2015_ccm_J1J2sq_spinGap}, where possible reasons for
  the difference in accuracy of the fits for small and large values of
  $\kappa$ were discussed.  Despite this difference in the quality of
  the extrapolations for the N\'{e}el and striped phases, the results
  in Fig.\ \ref{E_gap}(a) are clearly compatible with $\Delta$ being
  zero in the ranges $\kappa < \kappa_{c_{1}}$ and $\kappa >
  \kappa_{c_{2}}$ for the two quasiclassical GS phases with magnetic
  LRO, where the spin gap must be zero.

  Conversely, we see from Fig.\ \ref{E_gap}(a) that $\Delta > 0$ for a
  considerable range of values in the range $\kappa_{c_{1}} < \kappa <
  \kappa_{c_{2}}$ on both the N\'{e}el $(\kappa > \kappa_{c_{1}})$ and
  striped $(\kappa < \kappa_{c_{2}})$ sides of the region, over which
  the LSUB$n$ solutions with $n \leq 12$ persist, before their
  respective termination points.  The values of $\kappa$ at which the
  LSUB$\infty$ curves for $\Delta$ become nonzero are also completely
  compatible with the values for $\kappa_{c_{1}}$ and $\kappa_{c_{2}}$
  determined by us previously \cite{DJJF:2011_honeycomb}, at which the
  magnetic order parameter vanishes, $M \rightarrow 0$.  Our results
  are incompatible with the entire intermediate phase for
  $\kappa_{c_{1}} < \kappa < \kappa_{c_{2}}$ being a gapless spin
  liquid.  By contrast they provide supporting evidence to our earlier
  findings \cite{DJJF:2011_honeycomb} that the quantum paramagnetic
  phase in this regime has PVBC order.  On the other hand, our results
  for $\Delta$ would not, by themselves, rule out a gapped spin liquid in the
  intermediate regime.

  It is interesting to note that our results for $\Delta$ being
  nonzero over (at least part of) the intermediate regime for the
  present honeycomb model, are in contrast to the equivalent CCM
  results \cite{Richter:2015_ccm_J1J2sq_spinGap} for the corresponding
  intermediate quantum paramagnetic regime in the spin-$\frac{1}{2}$
  $J_{1}$--$J_{2}$ model on the square lattice.  For the latter case
  the extrapolated values of $\Delta$ over the entire parameter region
  accessible were found to be zero (or very close to zero).  For the
  spin-$\frac{1}{2}$ $J_{1}$--$J_{2}$ model on the square lattice the
  N\'{e}el order is found to melt at a comparable value of the
  frustration parameter, $\kappa \equiv J_{2}/J_{1}$, at
  $\kappa_{c_{1}} \approx 0.45$, with a paramagnetic state in the
  region $\kappa_{c_{1}} < \kappa < \kappa_{c_{2}} \approx 0.59$.  In
  this case the CCM based on the N\'{e}el state as model state can
  access the region $\kappa_{c_{1}} < \kappa \lesssim 0.49$ in LSUB$n$
  approximations with $n \leq 12$
  \cite{Richter:2015_ccm_J1J2sq_spinGap}, and in this region around
  the seemingly continuous transition at $\kappa_{c_{1}}$ the
  extrapolated CCM value of $\Delta$ is zero within computational
  accuracy.  A very recent, accurate, density-matrix renormalization
  group calculation \cite{Gong:2014_J1J2mod_sqLatt} also finds
  good evidence for a near-critical state in the region
  $\kappa_{c_{1}} < \kappa \lesssim 0.5$, with a very small gap for
  the finite-sized systems studied, and also for a gapped PVBC state
  in the remainder of the paramagnetic regime, $0.5 \lesssim \kappa <
  \kappa_{c_{2}}$.

  Finally, on the issue of the spin gap, we also present corresponding
  results in Fig.\ \ref{E_gap}(b) for the current model obtained by
  ED, for lattices containing $N=24, 32$ spins.  Clearly, neither
  lattice is large enough to show clearly the $N \rightarrow \infty$
  behavior of $\Delta=0$ over both quasiclassical regimes (i.e.,
  N\'{e}el and striped collinear).  Nevertheless, there is evidence
  (interestingly, more marked for the smaller lattice) of a gap
  opening up around $\kappa \approx 0.6$ (i.e., at $\kappa_{c_{2}}$)
  as $\kappa$ is decreased from above.  The steep decay around this
  value (seen most clearly in the $N=32$ data) is due to a level
  crossing in the triplet state.  While the ED data are clearly
  indicating the first-order transition at $\kappa = \kappa_{c_{2}}$,
  they are, unsurprisingly, quite smooth around the continuous
  transition at $\kappa=\kappa_{c_{1}}$.  Thus, while the ED results
  certainly do not contradict our CCM results, by themselves they are
  certainly much less predictive than the corresponding CCM results.
  Certainly, they do not, by themselves, contradict the appearance of
  a gap in the intermediate regime $\kappa_{c_{1}} < \kappa <
  \kappa_{c_{2}}$ with a sizable peak value, $\Delta \approx 0.4$, as
  shown by the CCM results.

  We turn now to our results for the low-energy parameters, $\rho_{s}$
  and $\chi$.  In view of the lower symmetries of the twisted and
  canted CCM model states used respectively for these parameters, as
  illustrated in Figs.\ \ref{pattern_sStiff} and
  \ref{pattern_M_ExtField}, the numbers $N_{f}$ of fundamental
  configurations at a given LSUB$n$ level of approximation are greater
  than those for the GS parameters $E/N$ and $M$ and for the spin gap
  $\Delta$.  For example, for the calculation of $\Delta$, $N_{f}
  \approx 10^{4}\, (2 \times 10^{5})$ for LSUB$n$ approximations based
  on the N\'{e}el model state with $n=10\, (12)$.  Corresponding values
  based on the striped model state are $N_{f} \approx 2.5 \times
  10^{4}\, (5 \times 10^{5})$ with $n=10\, (12)$.  By contrast, for the
  twisted model states used for the calculation of $\rho_{s}$, as
  shown in Fig.\ \ref{pattern_sStiff}, $N_{f} \approx 3.5 \times
  10^{5}$ at the LSUB10 level of approximation, and it becomes
  computationally infeasible to perform LSUB$n$ calculation for
  $\rho_{s}$ at higher truncation levels, $n \geq 12$.

\begin{figure}
\includegraphics[angle=270,width=8.5cm]{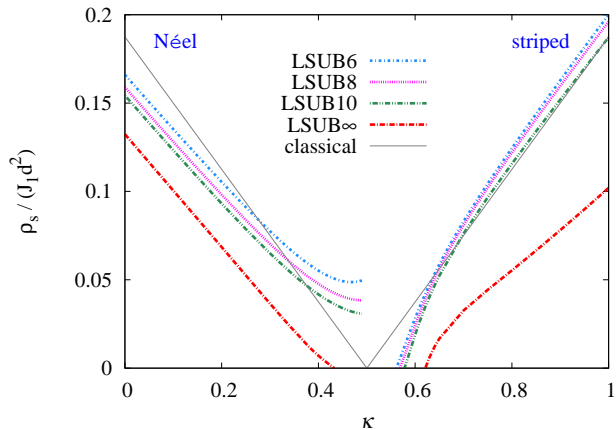}
\caption{(Color online) CCM results for the spin stiffness $\rho_{s}$
  (in units of $J_{1}d^{2}$) versus the frustration parameter $\kappa
  \equiv J_{2}/J_{1}$, for the spin-$\frac{1}{2}$
  $J_{1}$--$J_{2}$--$J_{3}$ model on the honeycomb lattice (with
  $J_{1}=1$, $J_{3}=J_{2}>0$).  We show results based on the N\'{e}el AFM
  and the striped AFM states as CCM model states.  The LSUB$n$ results
  are shown with $n=\{6,8,10\}$, together with the extrapolated
  LSUB$\infty$ results using Eq.\ (\ref{Eq_sstiff}) with this data
  set.  The classical results from Eqs.\ (\ref{sStiff_neel_classical})
  and (\ref{sStiff_stripe_classical}) are also shown for the value
  $s=\frac{1}{2}$.}
\label{spinstiff}
\end{figure}

Figure \ref{spinstiff} shows our ``raw'' CCM LSUB$n$ results with $n=\{6,8,10\}$ for $\rho_{s}$, together with the corresponding LSUB$\infty$ values using the extrapolation scheme of Eq.\ (\ref{Eq_sstiff}) together with this data set.
For comparison purposes, we also show in Fig.\ \ref{spinstiff} the
corresponding classical values obtained from Eqs.\
(\ref{sStiff_neel_classical}) and (\ref{sStiff_stripe_classical}) by
putting $J_{3} = J_{2} = \kappa J_{1}$ and $s=\frac{1}{2}$, viz.,
$\rho^{{\rm N\acute{e}el}}_{s;\,{\rm
    cl}}/(J_{1}d^{2})=\frac{3}{16}(1-2\kappa)$, and $\rho^{{\rm
    striped}}_{s;\,{\rm cl}}/(J_{1}d^{2})=\frac{3}{16}(-1+2\kappa)$.
We see clearly that the extrapolated values for $\rho_{s}$ for both
(the N\'{e}el and striped collinear) quasiclassical phases lie
somewhat lower than their classical counterparts for all values of the
frustration parameter $\kappa$.  Based on LSUB$n$ extrapolations of
$\rho_{s}$ with $n=\{6,8,10\}$, the critical values at which $\rho_{s}
\rightarrow 0$ on the N\'{e}el and striped sides of the phase
diagram, respectively, are $\kappa_{c_{1}} \approx 0.433$ and
$\kappa_{c_{2}} \approx 0.621$.  These may be compared with the
corresponding values in our earlier paper \cite{DJJF:2011_honeycomb},
based on LSUB$n$ extrapolations of the corresponding points where the
magnetic order parameter $M \rightarrow 0$, of $\kappa_{c_{1}} \approx
0.448$ and $\kappa_{c_{2}} \approx 0.601$ based on the same data set
$n=\{6,8,10\}$, and the presumably even more accurate values
$\kappa_{c_{1}} \approx 0.466$ and $\kappa_{c_{2}} \approx 0.601$
based on the larger data set $n=\{6,8,10,12\}$ available in this case.
Clearly, the totally independent estimates of the two QCPs from the
places where $M \rightarrow 0$ and $\rho_{s} \rightarrow 0$ are in
good agreement with one another, within very small errors arising from
the extrapolations.

In order to estimate the accuracy of our results for $\rho_{s}$
independently, we may compare with those of others for the special
case of a pure honeycomb-lattice HAF with NN interactions only, i.e.,
when $\kappa=0$.  Our LSUB$\infty$ extrapolated value using the
LSUB$n$ data set with $n=\{6,8,10\}$ is $\rho_{s}(\kappa=0) \approx
0.1324J_{1}d^{2}$.  To our knowledge the best available alternative
result for $\rho_{s}(\kappa=0)$ comes from a first-principles QMC
method using a highly efficient loop-cluster algorithm
\cite{Jiang:2008_honey,Jiang:2012_honey}.  The most accurate value
quoted by Jiang \cite{Jiang:2012_honey} is
$\bar{\rho}_{s}(\kappa=0)=0.1012(2)J_{1}$, equivalent to
$\rho_{s}(\kappa=0)=0.1315(3)J_{1}d^{2}$, which is in excellent
agreement with our own result.

\begin{figure}
\includegraphics[angle=270,width=8.5cm]{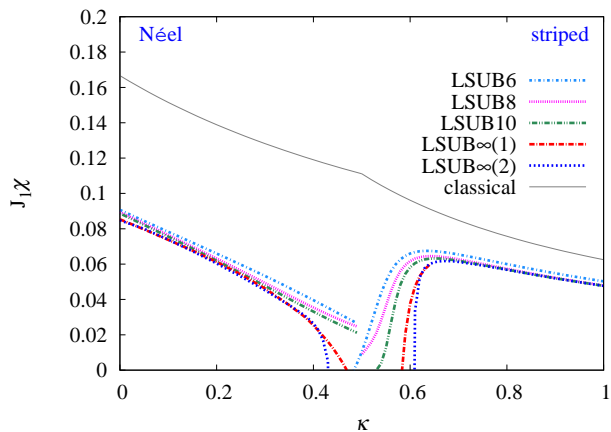}
\caption{(Color online) CCM results for the zero-field magnetic
  susceptibility (in units of $J_{1}^{-1}$) versus the frustration parameter $\kappa \equiv
  J_{2}/J_{1}$, for the spin-$\frac{1}{2}$
  $J_{1}$--$J_{2}$--$J_{3}$ model on the honeycomb lattice (with
  $J_{1}=1$, $J_{3}=J_{2}>0$).  We show results based on the N\'{e}el and
  the striped AFM states as CCM model states.  The LSUB$n$ results are
  shown with $n=\{6,8,10\}$, together with the extrapolated
  LSUB$\infty(1)$ and LSUB$\infty(2)$ results using Eqs.\ (\ref{Eq_X})
  and (\ref{Eq_X_nu}), respectively, with this data set.  The
  classical results from Eqs.\ (\ref{chi_neel_classical}) and
  (\ref{chi_stripe_classical}) are also shown for the value
  $s=\frac{1}{2}$.}
\label{M_Xcpty}
\end{figure}

Figure \ref{M_Xcpty} shows our corresponding results for the
zero-field magnetic susceptibility $\chi$.
For the striped collinear state used as the CCM model state, the
number $N_{f}$ of fundamental configurations at the LSUB10 level is
$N_{f} \approx 3.5 \times 10^{5}$, and once again, just as for
$\rho_{s}$ it is computationally infeasible to perform LSUB$n$
calculations for this phase with $n \geq 12$.  By contrast, for the
N\'{e}el state used as our CCM model state, $N_{f} \approx 6 \times
10^{4}$ at the LSUB10 level and $N_{f} \approx 1.1 \times 10^{6}$ at
the LSUB12 level.  In order to be consistent with our treatment of the
two quasiclassical phases, we show the ``raw'' LSUB$n$ results in
Fig.\ \ref{M_Xcpty} with $n=\{6,8,10\}$ and corresponding LSUB$\infty$
extrapolations based on this data set.  However, on the N\'{e}el side
we have also performed an LSUB12 calculation for the unfrustrated
limiting case $\kappa=0$.  Once again, for comparison purposes, we
also show in Fig.\ \ref{M_Xcpty} the corresponding classical values
obtained from Eqs.\ (\ref{chi_neel_classical}) and
(\ref{chi_stripe_classical}) by putting $J_{3}=J_{2}=\kappa J_{1}$ and
$s=\frac{1}{2}$, viz., $\chi^{\rm N\acute{e}el}_{{\rm
    cl}}=1/[6J_{1}(1+\kappa)]$, and $\chi^{{\rm striped}}_{{\rm
    cl}}=1/[2J_{1}(1+7\kappa)]$.

Once again we see that quantum fluctuations act to reduce the value of
$\chi$ considerably from its classical value in both quasiclassical
AFM phases.  Figure \ref{M_Xcpty} also shows that the two
extrapolation schemes of Eqs.\ (\ref{Eq_X}) and (\ref{Eq_X_nu}) give
results in very close agreement with one another on both sides of the
phase diagram, except in very narrow regions close to the two QCPs at
$\kappa=\kappa_{c_{1}}$ and $\kappa=\kappa_{c_{2}}$.  Perhaps the most
important feature of Fig.\ \ref{M_Xcpty} is that, unlike in the
classical case where $\chi_{{\rm cl}}$ takes the (nonzero) value
$\frac{1}{9}$ at the phase transition point $\kappa_{{\rm
    cl}}=\frac{1}{2}$, in the quantum $s=\frac{1}{2}$ case $\chi$ now
vanishes at the two QCPs, which is a very clear indication of a spin
gap opening up at these points
\cite{Mila:2000_M-Xcpty_spinGap,Bernu:2015_M-Xcpty_spinGap}.

We show in Fig.\ \ref{M_Xcpty} extrapolated LSUB$\infty$ results based
on the schemes of both Eqs.\ (\ref{Eq_X}) and (\ref{Eq_X_nu}).  The
two sets of results shown are seen to be in excellent agreement with
each other except in very narrow regimes close to the two QCPs at
$\kappa_{c_{1}}$ and $\kappa_{c_{2}}$.  Based on the scheme of Eq.\
(\ref{Eq_X}), the two critical values obtained from extrapolating the
LSUB$n$ values with $n=\{6,8,10\}$ are $\kappa_{c_{1}} \approx 0.469$
and $\kappa_{c_{2}} \approx 0.583$.  The corresponding values from
using scheme of Eq.\ (\ref{Eq_X_nu}) are $\kappa_{c_{1}} \approx
0.430$ and $\kappa_{c_{2}} \approx 0.609$.  While both sets of results
agree quite well with the corresponding results cited above of
$\kappa_{c_{1}} \approx 0.448$ and $\kappa_{c_{2}} \approx 0.601$ from
our earlier work on extrapolations for the magnetic order parameter of
corresponding LSUB$n$ results with $n=\{6,8,10\}$, there is clearly
more uncertainty in the critical values obtained from the zero-field
magnetic susceptibility.

As an aside here, we remark that the shapes (e.g., the slopes) of the
respective extrapolated (LSUB$\infty$) curves for $\rho_{s}$ and
$\chi$ in Figs.\ \ref{spinstiff} and \ref{M_Xcpty} appear to vary
significantly near where they vanish at the QCP $\kappa_{c_{1}}$, and
also, but to a lesser extent, at the QCP $\kappa_{c_{2}}$.  Clearly,
in turn, this would imply different values for the critical indices
for the two low-energy parameters at $\kappa_{c_{1}}$, in particular.
However, we remain cautious about putting too much weight on such an
interpretation, since it is precisely in the regions very close to the
QCPs where the extrapolations become most demanding.  While we are
rather confident of the calculated values for $\kappa_{c_{1}}$ and
$\kappa_{c_{2}}$ themselves, within errors we can estimate (and also
see the further discussion below in Sec.\ \ref{summary_sec}), the
actual detailed shapes of the extrapolated curves (including, e.g.,
their slopes) at the QCPs is more open to doubt and the errors are
more difficult to quantify.  A close comparison of the raw LSUB$n$
families of curves and their respective LSUB$\infty$ extrapolations in
Figs.\ \ref{spinstiff} and \ref{M_Xcpty} shows that the slopes of the
raw and extrapolated curves very near to a supposedly continuous
transition like that at $\kappa_{c_{1}}$ can differ appreciably.  This
difference is less marked at a first-order transition like that at
$\kappa_{c_{2}}$, although still present to a lesser degree.  For
these reasons we are reluctant, with our present methodology and
accuracy, to make any claims to be able to calculate, with any
quantitative degree of accuracy, the critical indices for the
vanishing of $\rho_{s}$ and $\chi$ at the two QCPs.  Any such
interpretation drawn from our results about the critical indices being
different for $\rho_{s}$ and $\chi$ at $\kappa_{c_{1}}$ especially,
should be regarded as suggestive at best, in our opinion.

It is also of interest to compare our results for $\chi$ for the
limiting case $\kappa=0$ of a pure honeycomb HAF with NN interactions
only.  Our extrapolated LSUB$\infty$ results based on Eq.\
(\ref{Eq_X}) are $\chi(\kappa=0)=0.0847(4)/J_{1}$, using LSUB$n$ data
points $n=\{6,8,10,12\}$, where the quoted errors is purely that
associated with the fit.  Corresponding extrapolations based on
LSUB$n$ results with $n=\{6,8,10\}$ and $n=\{8,10,12\}$ are,
respectively, $\chi(\kappa=0) \approx 0.0845/J_{1}$ and
$\chi(\kappa=0) \approx 0.0837/J_{1}$.  Other estimates are
$\chi(\kappa=0) \approx 0.0756(10)/J_{1}$ from a linked-cluster SE
analysis \cite{Oitmaa:1992_honey}; $\chi(\kappa=0) \approx
0.1667/J_{1}$ and $\chi(\kappa=0) \approx 0.0456/J_{1}$ from SWT at
leading (classical) order and next-to-leading order [viz., with
$O(1/s)$ corrections included], respectively \cite{Oitmaa:1992_honey};
and $\chi(\kappa=0)=0.0666/J_{1}$ from SBMFT
\cite{Mattsson:1994_honey}.  The most accurate estimates for this
unfrustrated limit undoubtedly come from QMC calculations.  For
example, L\"{o}w \cite{Low:2009_honey} used a continuous Euclidean
time QMC algorithm to find a value
$\hat{\chi}(\kappa=0)=0.05188(8)/J_{1}$.  Presumably, in the QMC calculations of the magnetic susceptibility the value $\hat{\chi}$ obtained is an average over all directions of the applied field, since the ground state is calculated for a finite system, i.e., with no breaking of the rotational symmetry.  By contrast, in the CCM calculations we start with a symmetry-broken state and apply the field perpendicular to the axis of the order parameter, to obtain $\chi_{\perp}\,(\equiv \chi)$ directly.  Thus, $\hat{\chi}=\frac{1}{3}(\chi_{\parallel}+2\chi_{\perp})$, where $\chi_{\parallel}\,(=0)$ is the parallel component of the magnetic susceptibility.  Hence, we have $\chi=\frac{3}{2}\hat{\chi}$, and the QMC result of L\"{o}w is equivalent to $\chi(\kappa=0)=0.0778(1)/J_{1}$.  
Another QMC estimate, using a loop-cluster algorithm by Jiang
\cite{Jiang:2012_honey}, can also be quoted.  While Jiang does not
quote a result for $\chi$ directly, he does provide the result $\hbar
c(\kappa=0)=1.2905(8)J_{1}d$.  This may be combined with his result for
$\rho_{s}$ cited above to calculate $\chi=\rho_{s}/(\hbar c)^{2}$,
yielding the value $\chi(\kappa=0)=0.0789(3)/J_{1}$.  Clearly, our own
best CCM estimate, $\chi(\kappa=0)=0.084(2)/J_{1}$, lies slightly
higher than those two QMC estimates.

\begin{figure}[!t]
\includegraphics[angle=270,width=8.5cm]{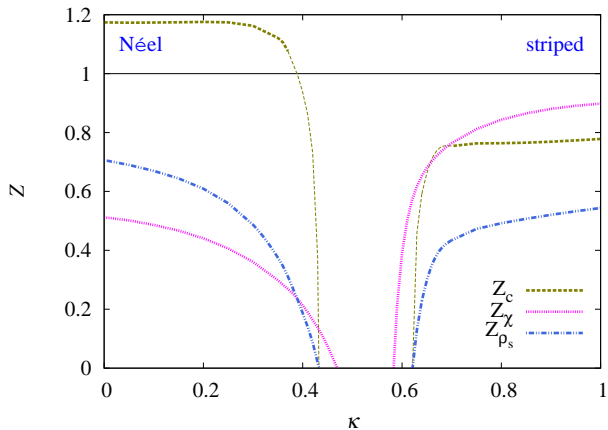}
\caption{(Color online) CCM results for the three renormalization
  constants corresponding to the spin stiffness $\rho_{s}$, the
  zero-field magnetic susceptibility $\chi$, and the spin-wave
  velocity, $c$, of the spin-$\frac{1}{2}$ $J_{1}$--$J_{2}$--$J_{3}$
  model on the honeycomb lattice with $J_{1}> 0$, in the case
  $J_{3}=J_{2}=\kappa J_{1}$.  Results for $Z_{c}$ are unreliable in the regions shown by thin lines, close to the critical points (and see text for details).}
\label{Z_M_Xcpty_sStiff}
\end{figure}

\begin{table*}[!t]
  \caption{The low-energy parameters for the spin-$\frac{1}{2}$ HAF on the honeycomb lattice with lattice spacing $d$ and with exchange interactions between NN pairs only, all with equal strength $J_{1}>0$.  The classical values are compared with our CCM results, and those using the alternative techniques of linked-cluster SE \cite{Oitmaa:1992_honey} and two different QMC algorithms \cite{Low:2009_honey,Jiang:2012_honey}.}
\vskip0.2cm
\begin{tabular}{cccccc} \hline\hline
Parameter & Classical & CCM & SE                              & QMC & QMC \\ 
          & value     &     & (Ref.\ \cite{Oitmaa:1992_honey}) & (Ref.\ \cite{Low:2009_honey}) & (Ref.\ \cite{Jiang:2012_honey}) \\ \hline
$E/(NJ_{1})$ & -0.375 & -0.54466(2) & -0.5443(3) & -0.54455(20) & \\
$M$          & 0.5    & 0.2714(10) & 0.266(9)   & 0.2681(8) & 0.26882(3) \\
$\rho_{s}/(J_{1}d^{2})$ & 0.1875 & 0.1324(5) & & & 0.1315(3) \\
$J_{1}\chi$ & 0.1667 & 0.084(2) & 0.0756(10) & 0.07782(12) & \\
$\hbar c/(J_{1}d)$ & 1.0607 & 1.255(15) & & & 1.2905(8) \\ \hline\hline
\end{tabular} 
\label{table_summary} 
\end{table*}

It is also convenient to express the low-energy parameters in
terms of a multiplicative renormalization constant $Z$ with respect to
the corresponding classical result.  Thus, we define $Z_{\rho_{s}}$
and $Z_{\chi}$ as
\begin{equation}
Z_{\rho_{s}}\equiv \rho_{s}/\rho_{s;\,{\rm cl}}\,, \qquad Z_{\chi} \equiv \chi/\chi_{{\rm cl}}\,.  \label{eq_Z-rho_Z-chi}
\end{equation}
Since the spin-wave velocity is given by $\hbar c =
\sqrt{\rho_{s}/\chi}$, its corresponding renormalization constant is
\begin{equation}
Z_{c} = \sqrt{Z_{\rho_{s}}/Z_{\chi}}\,.   \label{eq_Z-c}
\end{equation}
Our extrapolated LSUB$\infty$ results for the three renormalization
constants of Eqs.\ (\ref{eq_Z-rho_Z-chi}) and (\ref{eq_Z-c}) are shown
in Fig.\ \ref{Z_M_Xcpty_sStiff}, where we employ the LSUB$\infty(1)$
results for $\chi$ obtained from the extrapolation scheme of Eq.\
(\ref{Eq_X}).

We should point out that the seeming vanishing of the Goldstone
spin-wave velocity at both critical points $\kappa_{c_{1}}$ and
$\kappa_{c_{2}}$, as shown in Fig.\ \ref{Z_M_Xcpty_sStiff}, is
entirely an artefact of our calculational scheme.  Thus, $Z_{c}$ has
been calculated here indirectly via Eq.\ (\ref{eq_Z-c}) using our
(separately) extrapolated results for $\rho_{s}$ and $\chi$.  In
reality both of these parameters should go to zero at the same points
(at least for $\chi$ at a transition at which a spin gap opens).
However, as we have noted, our calculated results for $\rho_{s}$ and
$\chi$ are completely independent of one another, and as a consequence
the critical values at which both parameters vanish are not
constrained to be the same.  This is both a strength and a weakness of
the methodology.  The main advantage is that it provides an inbuilt
error estimate for the accuracy of our calculated values of
$\kappa_{c_{1}}$ and $\kappa_{c_{2}}$.  However, it is also clearly a
disadvantage when we wish to take ratios, as in Eq.\ (\ref{eq_Z-c}),
in regimes close to the QCPs.  Thus, if $\rho_{s}$ and $\chi$ go to
zero at slightly different values of $\kappa$, it is guaranteed that
the spin-wave velocity calculated in terms of them will, quite
artificially, approach either zero or infinity near the actual QCP.
We have thus indicated in Fig.\ \ref{Z_M_Xcpty_sStiff}, by thinner
portions of the curves for $Z_{c}$, those regions near the two QCPs
where the results for $Z_{c}$ are correspondingly unreliable.

Finally, we collect in Table \ref{table_summary} our CCM results for
the full set of low-energy parameters for the unfrustrated
($\kappa=0$) limiting case of a pure spin-$\frac{1}{2}$ HAF on the
honeycomb lattice with NN interactions only (of strength $J_{1}>0$).
We compare our results there with values obtained from a
linked-cluster SE analysis \cite{Oitmaa:1992_honey}, and two different
QMC analyses \cite{Low:2009_honey,Jiang:2012_honey}, which are
expected to be very accurate in this case, where the infamous
minus-sign problem is absent.  There is no reason to expect that the
evident accuracy of our results in this limit will be any lower over
the entire range of values of $\kappa$ accessible to us, by strong
contrast with QMC techniques, which degrade significantly in the
presence of frustration.

\section{SUMMARY AND CONCLUSIONS}
\label{summary_sec}
In this paper we have continued our prior investigation
\cite{DJJF:2011_honeycomb} of the spin-$\frac{1}{2}$
$J_{1}$--$J_{2}$--$J_{3}$ HAF model on the honeycomb lattice, along
the line $J_{3}=J_{2}=\kappa J_{1}$, with $J_{1}>0$, that includes the
point of maximum classical frustration at $\kappa = \kappa_{{\rm cl}}
= \frac{1}{2}$.  Just as in our prior work we have used the CCM based
on both the N\'{e}el and collinear striped AFM states as reference
states, with respect to which we have included quantum fluctuations in
the fully consistent LSUB$n$ truncation scheme.  We have carried out
calculations to high orders in the truncation index $n$ (typically
with $n \leq 10$, but in some cases also with $n \leq 12$).  The only
approximation made has been to extrapolate in the truncation index
$n$ to the exact ($n \rightarrow \infty$) LSUB$\infty$ limit.

We have now calculated a complete set of low-energy parameters ($E/N,
M, \rho_{s}, \chi$, and $c$) for the model, from which we obtain
independent pieces of evidence that provide a clear and consistent
description of its $T=0$ quantum phase diagram.  In particular, we
find compelling evidence that the single phase transition in the
classical ($s \rightarrow \infty$) version of the model at
$\kappa_{{\rm cl}}=\frac{1}{2}$ is split in the $s=\frac{1}{2}$ model
into two quantum phase transitions at $\kappa_{c_{1}}$ and
$\kappa_{c_{2}}$.  In our prior work \cite{DJJF:2011_honeycomb} these
two QCPs were identified independently by the points at which both the
magnetic order parameter $M$ and the inverse of the susceptibility
coefficient against the formation of a state with PVBC order,
$1/\chi_{p}$, vanish.  To those estimates we now add two more, based
on the points at which both $\rho_{s}$ and $\chi$ vanish.  The
collected results are displayed together in Table \ref{table_critPts},
in which the results for $M$ and $1/\chi_{p}$ are obtained
\cite{DJJF:2011_honeycomb} from LSUB$\infty$ extrapolation based on
LSUB$n$ data points with $n=\{6,8,10,12\}$, while those for $\rho_{s}$
and $\chi$ are obtained from data points with $n=\{6,8,10\}$.  All of
our results are fully consistent with values $\kappa_{c_{1}}=0.45(2)$
and $\kappa_{c_{2}}=0.60(2)$.
\begin{table}[!b]
  \caption{Values of the two quantum critical points $\kappa_{c_{1}}$ and $\kappa_{c_{2}}$ of the spin-$\frac{1}{2}$ $J_{1}$--$J_{2}$--$J_{3}$ HAF model on the honeycomb lattice, along the line $J_{3}=J_{2}=\kappa J_{1}$  with ($J_{1}>0$), as obtained from the vanishing points of the magnetic order parameter $M$, the spin stiffness $\rho_{s}$, the uniform magnetic susceptibility, and the inverse of the susceptibility coefficient of the system against the formation of PVBC order $1/\chi_{p}$, all evaluated by the extrapolation of CCM LSUB$n$ results.}
\vskip0.2cm
\begin{tabular}{ccc} \hline\hline
Parameter ($\rightarrow 0$) & $\kappa_{c_{1}}$ & $\kappa_{c_{2}}$  \\ \hline
$M$ (see Ref.\ \cite{DJJF:2011_honeycomb}) & 0.466    & 0.601  \\
$\rho_{s}$ (this work) & 0.433 & 0.621 \\
$\chi$ (this work): LSUB$\infty(1)$ $^{a}$ & 0.469 & 0.583  \\
$\chi$ (this work): LSUB$\infty(2)$ $^{b}$ & 0.430 & 0.609  \\ \hline
$1/\chi_{p}$ (see Ref.\ \cite{DJJF:2011_honeycomb}) & 0.473 & 0.586 \\ \hline\hline
\end{tabular} 
\vskip0.3cm
\protect 
$^{a}$Using the extrapolation scheme of Eq.\ (\ref{Eq_X}). \\
$^{b}$Using the extrapolation scheme of Eq.\ (\ref{Eq_X_nu}).  
\label{table_critPts} 
\end{table}

Based on the shape of the curves for $M$ and $1/\chi_{p}$ as functions
of $\kappa$ we suggested in our earlier work that the QCP at
$\kappa_{c_{1}}$ between the state with N\'{e}el order (for $\kappa <
\kappa_{c_{1}}$) and the paramagnetic intermediate state was of
continuous type, while that at $\kappa_{c_{2}}$ between the state with
collinear striped order (for $\kappa > \kappa_{c_{2}}$) and the
intermediate state is of first-order type.  The new evidence, based on
$\rho_{s}$ and $\chi$, shown in Figs.\ \ref{spinstiff} and
\ref{M_Xcpty} respectively, is completely consistent with this
interpretation, as indeed is also the evidence based on the curves for
the spin gap $\Delta$ shown in Figs.\ \ref{E_gap}(a) and
\ref{E_gap}(b).

Since at a QCP the quantum fluctuations present in a system are
sufficiently strong to make the system infinitely susceptibility to
multiple forms of order, the vanishing of $1/\chi_{p}$ at
$\kappa_{c_{1}}$ and $\kappa_{c_{2}}$ cannot be taken as strong
evidence that PVBC order is present over the entire intermediate
regime, $\kappa_{c_{1}} < \kappa < \kappa_{c_{2}}$.  However, just as
we could in this work find values of $\Delta$ into this regime in
regions around both QCPs, so in our earlier work
\cite{DJJF:2011_honeycomb} could we calculate $\chi_{p}$ in similar
regions.  The shape of the curves for $1/\chi_{p}$ as a
function of $\kappa$ was consistent with $1/\chi_{p}$ vanishing in
those regions.  It was on this evidence that we made the tentative
conclusion that the entire region $\kappa_{c_{1}} < \kappa <
\kappa_{c_{2}}$ contained a quantum phase with PVBC order.

The results of the present paper provide strong support for such an
interpretation.  Firstly, the vanishing of $\chi$ at $\kappa_{c_{1}}$
and $\kappa_{c_{2}}$ provides compelling evidence for a gapped state
opening up at these points.  Secondly, our results displayed in Fig.\
\ref{E_gap} provide positive and conclusive evidence of a gapped state
over a considerable range of the intermediate regime, which is
accessible using both quasiclassical AFM states as CCM model states.
The values at which $\Delta$ becomes nonzero are also wholly
consistent with the values for $\kappa_{c_{1}}$ and $\kappa_{c_{2}}$
cited above, as obtained from the GS low-energy parameters.

We note that Goldstone's theorem implies that any state that breaks
spin-rotational symmetry must have a vanishing gap.  Thus, the
non-vanishing of the triplet gap $\Delta$ in the intermediate phase
completely rules out the possibility of {\it any} type of magnetic
order being present in this regime, not only the N\'{e}el and striped
forms considered explicitly here.  Similarly, the fact that $\Delta
\neq 0$ in the intermediate phase also precludes other more exotic
forms of order that break SU(2) symmetry.  Examples include
spin-nematic states, which break SU(2) symmetry while preserving
translational and time-reversal symmetries.

Finally, we note that it might also be interesting in future work to
calculate the singlet excitation gap within the disordered regime, in
order to compare it with the triplet gap calculated here.  While the
singlet gap can certainly also be calculated within the CCM framework,
the calculations are more challenging, since the excited state now
lies in the same sector as the ground state, viz., with $S^{z}_{T} =
0$, where $S^{z}_{T} \equiv \sum^{N}_{i=1} s^{z}_{i}$.  A possible
motivation for doing so would be to compare with the corresponding
results for the spin-$\frac{1}{2}$ $J_{1}$--$J_{2}$ HAF on the square
lattice, the phase diagram for which is qualitatively similar to that
for the spin-$\frac{1}{2}$ $J_{1}$--$J_{2}$--$J_{3}$ Heisenberg model
on the honeycomb lattice, in the case $J_{3} = J_{2}$ considered here.

For example, a recent highly accurate density-matrix renormalization
group (DMRG) simulation of the spin-$\frac{1}{2}$ $J_{1}$--$J_{2}$
Heisenberg model on the square lattice \cite{Jiang:2012} found that
the singlet gap remains consistently below the triplet gap over the
intermediate disordered regime in this case.  The authors took this
finding as an indication of the formation of short-range singlets in
the intermediate phase.  In turn, this would be consistent with the
intermediate phase being a spin liquid or one with only weak
valence-bond crystalline (VBC) order.  By contrast, a phase with
stronger VBC order would be expected to have a triplon excitation as
the lowest-energy excited state, since such a state corresponds to the
breaking of only one singlet bond, compared to a singlet excitation
that requires the breaking of two singlet bonds.  If our conclusion
for the present case of a spin-$\frac{1}{2}$ $J_{1}$--$J_{2}$--$J_{3}$
Heisenberg model on the honeycomb lattice (with $J_{3} = J_{2}$), that
the intermediate phase has PVBC order, is correct, we would then expect the
singlet excitation gap to lie higher than the triplet gap, by contrast
with the above DMRG findings for the spin-$\frac{1}{2}$ HAF on the
square lattice.

In conclusion, the present paper revisits a model to which the CCM had
previously been applied, with the joint aims (and outcomes) to improve
the conceptual framework and to yield new physics, particularly with
regard to the nature of the intermediate phase.  Thus, the calculation
of a complete set of low-energy parameters for the model within a
single, unified, and consistent theoretical CCM framework, has not
only given more detailed information about each of the ordered
magnetic phases and more accurate values for their boundaries with the
intermediate (non-classical) disordered phase, but has also opened the
possibility for a full (quantitative) $\chi$PT treatment of the model.
Similarly, the CCM calculation of the triplet gap $\Delta$ has now
definitively ruled out the intermediate phase from having any form of
order that breaks SU(2) symmetry, including such exotic states as spin
nematics.

\section*{ACKNOWLEDGMENTS}
We thank the University of Minnesota Supercomputing Institute for the
grant of supercomputing facilities.  One of us (RFB) gratefully acknowledges the Leverhulme Trust for the award of an Emeritus Fellowship (EM-2015-07).  We also thank D.~J.~J.~Farnell for fruitful discussions in the early stages of this work.

\bibliographystyle{apsrev4-1}
\bibliography{bib_general}

\end{document}